\documentclass[useAMS,usenatbib]{mn2e}

\usepackage{graphicx}
\usepackage{amsmath}
\usepackage{amssymb}
\usepackage{color}
\usepackage{subfig}
\usepackage[breaklinks,colorlinks,citecolor=blue,linkcolor=red]{hyperref} 
\usepackage[all]{hypcap}

\newcommand\msun{\, \rm M_\odot}
\newcommand\rsun{\, \rm R_\odot}
\newcommand\km{\, \rm km}
\newcommand\kms{\, \rm km\,s^{-1}}

\newcommand\gyr{{\, \rm Gyr}}

\newcommand\eout{{e_{\rm out}}}
\newcommand\ein{{e_{\rm in}}}
\newcommand\aout{{a_{\rm out}}}
\newcommand\ain{{a_{\rm in}}}
\newcommand\amax{{a_{\rm 3, max}}}
\newcommand\mbh{{m_{\rm SBH}}}
\newcommand\mms{{m_{\rm MS}}}
\newcommand\mns{{m_{\rm NS}}}
\newcommand\sigbh{{\sigma_{\rm SBH}}}

\newcommand\vk{{v_\mathrm{k}}}

\title[TDEs in triples]{Tidal disruption events onto stellar black holes in triples}
\author[G. Fragione et al.]{  \parbox{\textwidth}{Giacomo Fragione$^{1,2,3}$\thanks{E-mail: giacomo.fragione@northwestern.edu}, Nathan W. C. Leigh$^{4,5}$, Rosalba Perna$^{6,7}$, Bence Kocsis$^{8}$ \vspace*{0.3cm}}\\
$^1$Racah Institute for Physics, The Hebrew University, Jerusalem 91904, Israel\\
$^2$Department of Physics \& Astronomy, Northwestern University, Evanston, IL 60202, USA\\
$^3$Center for Interdisciplinary Exploration \& Research in Astrophysics (CIERA), Evanston, IL 60202, USA\\
$^4$Departamento de Astronom\'ia, Facultad de Ciencias F\'isicas y Matem\'aticas, Universidad de Concepci\'on, Concepci\'on, Chile\\
$^5$Department of Astrophysics, American Museum of Natural History, New York, NY 10024, USA\\
$^6$Department of Physics and Astronomy, Stony Brook University, Stony Brook, NY 11794-3800, USA\\
$^7$Center for Computational Astrophysics, Flatiron Institute,  New York, NY 10010, USA\\
$^8$Institute of Physics, E\"{o}tv\"{o}s University, P\'azm\'{a}ny P. s. 1/A, Budapest, 1117, Hungary}

\begin{document}

\maketitle

\begin{abstract}
Stars passing too close to a black hole can produce tidal disruption events (TDEs), when the tidal force across the star exceeds the gravitational force that binds it. TDEs have usually been discussed in relation to massive black holes that reside in the centres of galaxies or lurk in star clusters. We investigate the possibility that triple stars hosting a stellar black hole (SBH) may be sources of TDEs. We start from a triple system made up of three main sequence (MS) stars and model the supernova (SN) kick event that led to the production of an inner binary comprised of a SBH. We evolve these triples with a high precision $N$-body code and study their TDEs as a result of Kozai-Lidov oscillations. We explore a variety of distributions of natal kicks imparted during the SN event, various maximum initial separations for the triples, and different distributions of eccentricities. We show that the main parameter that governs the properties of the SBH-MS binaries which produce a TDE in triples is the mean velocity of the natal kick distribution. Smaller $\sigma$'s lead to larger inner and outer semi-major axes of the systems that undergo a TDE, smaller SBH masses, and longer timescales. We find that the fraction of systems that produce a TDE is roughly independent of the initial conditions, while estimate a TDE rate of $2.1\times 10^{-4}-4.7 \ \mathrm{yr}^{-1}$, depending on the prescriptions for the SBH natal kicks. This rate is almost comparable to the expected TDE rate for massive black holes.
\end{abstract}


\section{Introduction}

Stars passing too close to a black hole can produce tidal disruption events (TDEs) when the tidal force across the star exceeds the gravitational force that binds it (see e.g. \citealt{stone2019} for a recent review). The resulting stellar debris can produce an electromagnetic flare and can be used as a powerful instrument to probe the presence of quiescent massive black holes, which would otherwise remain dark \citep{dorazio2019}. TDEs are usually discussed in the context of galactic nuclei, where the numerous 2-body interactions of stars surrounding a supermassive black hole (SMBH) drive some of them onto plunging orbits, which result in TDE events \citep{alex2017}. The rate of TDEs due to SMBHs in galactic nuclei is highly uncertain, and could be enhanced by a secondary black hole \citep{fraglei18}. Both observational and theoretical estimates are within the range $10^{-5}-10^{-4}$ yr$^{-1}$ per galaxy \citep{stone2016,alex2017,vanvelzen2018}. 

TDEs have also been studied in relation to the elusive intermediate-mass black holes (IMBHs) \citep{baumgardt2004,rosswog2008,rosswog2009}. The discovery of the tidal consumption of a star passing in the vicinity of an IMBH may provide a definitive proof for their existence. TDEs onto IMBHs can take place in galactic nuclei, where the rate of TDEs can be as high as $\sim 10^{-4}$--$10^{-2}$ yr$^{-1}$ on $\sim$ few Myr timescales. Another source of IMBH TDEs are globular clusters. Recent calculations by \citet{fraglgk2018} have computed a typical rate $\sim 10^{-5}$--$10^{-3}$ yr$^{-1}$. \citet{lin18} recently observed a TDE-like event consistent with an IMBH in an off-centre star cluster, at a distance of $\sim 12.5$ kpc from the centre of the host galaxy. 

Recently, TDEs have also been discussed in the context of stellar black holes (SBHs). \citet{rastello2019} showed that binary SBHs can trigger TDEs in open clusters and estimated a typical rate of $\sim 0.3$--$3\times 10^{-6}$ yr$^{-1}$. \citet{lopez2018} showed that the tidal disruption of a star by a SBH binary in a star cluster can alter the intrinsic spins of the two SBHs. \citet{sams2019} proposed that the same events can be used to constrain the orbital period distribution and, as a consequence, the dynamical mechanisms that eventually drive these binaries to merge. Recently, \citet{Kremer2019} used a wide range of N-body simulations of globular clusters and estimated a rate of $3\times 10^{-6}$ Gpc$^{-3}$ yr$^{-1}$ for these events.

Bound stellar multiples are not rare. Both massive (O, B, A type) and near solar mass stars (F, G, K type) have been shown to reside in triples or higher order multiplicities, with a relative fraction that can be as high as $\sim 20\%$--$30\%$ \citep{duq91,ragh10,sa2013AA,duns2015,sana2017,jim2019}. Most recent studies of transient events associated with bound multiple systems focused on determining the SBH and neutron star (NS) merger rates \citep{ant17,sil17,fragk2019,frloeb2019,liu2019,Fragione2019b} and the double white dwarf merger rate \citep{katz2012,fang2018,hamers2018,toonen2018}.

In this paper we investigate the possibility that isolated triple stars in galaxies are sources of TDEs, some of which can take place off-center from the nucleus of the host galaxy. We consider triples comprised of an inner binary consisting of a SBH and a main sequence star (MS), which eventually produce a TDE as a result of Kozai-Lidov (KL) oscillations. We start from the MS progenitors of the SBH and model the supernova (SN) event that leads to the formation of these triples under study. We adopt different prescriptions for the natal kick velocities that are imparted by SN events. We quantify how the probability of a TDE depends on the initial conditions, and determine the parameter distributions of merging systems relative to the initial distributions.

The paper is organized as follows. In Section~\ref{sect:supern}, we discuss the SN mechanism in the context of triple stars, and the KL mechanism. In Section~\ref{sect:results}, we present our numerical methods to determine the rate of SBH TDEs in triples, and discuss the parameters of the merging systems. The implications for  possible  electromagnetic counterparts are presented in Section~\ref{sect:em}.  Finally, in Section~\ref{sect:conc}, we discuss the implications of our findings for the current state of the literature and draw our conclusions.

\section{Supernovae in triples and the Kozai-Lidov mechanism}
\label{sect:supern}

\begin{figure*} 
\centering
\includegraphics[scale=0.55]{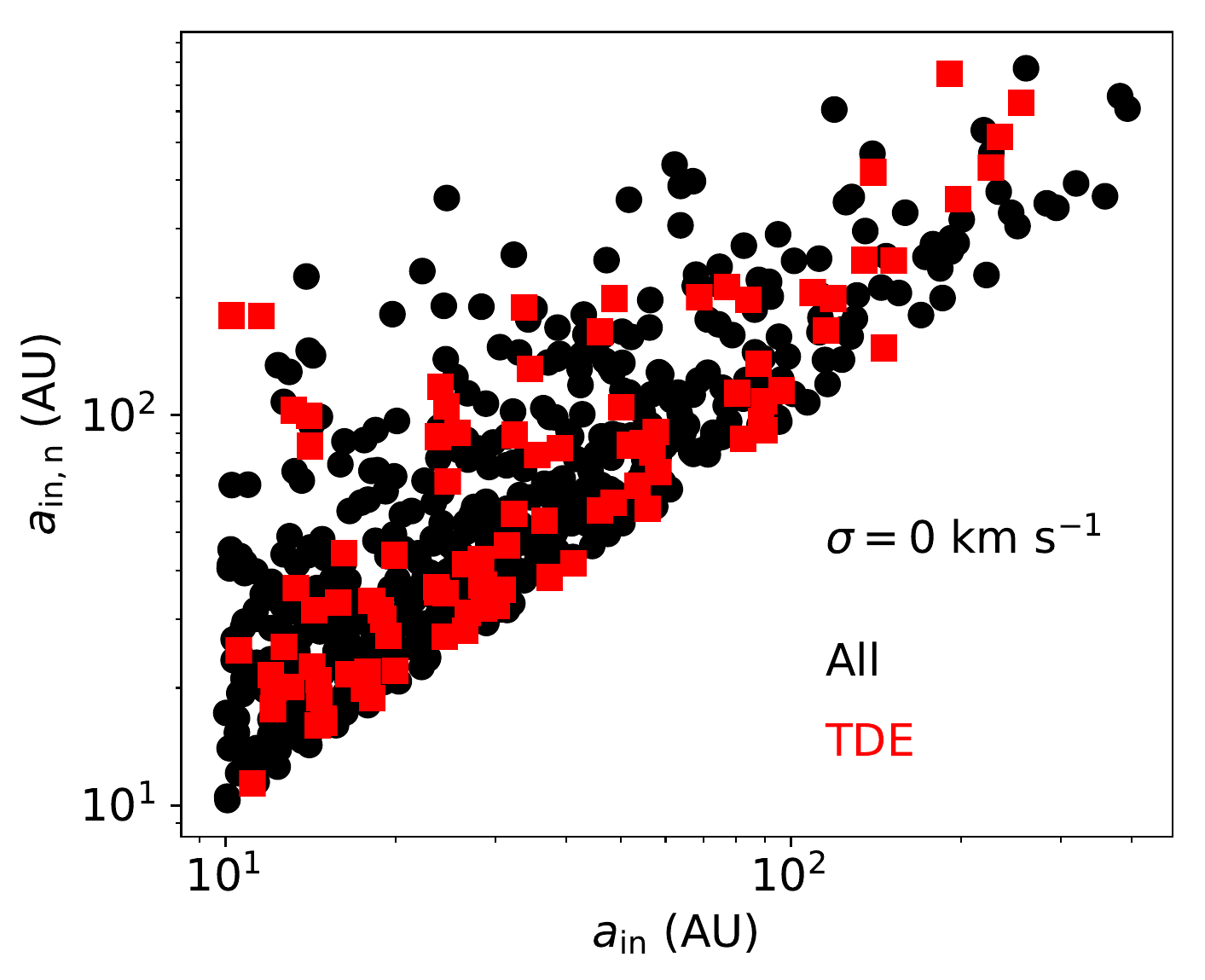}
\hspace{0.5cm}
\includegraphics[scale=0.55]{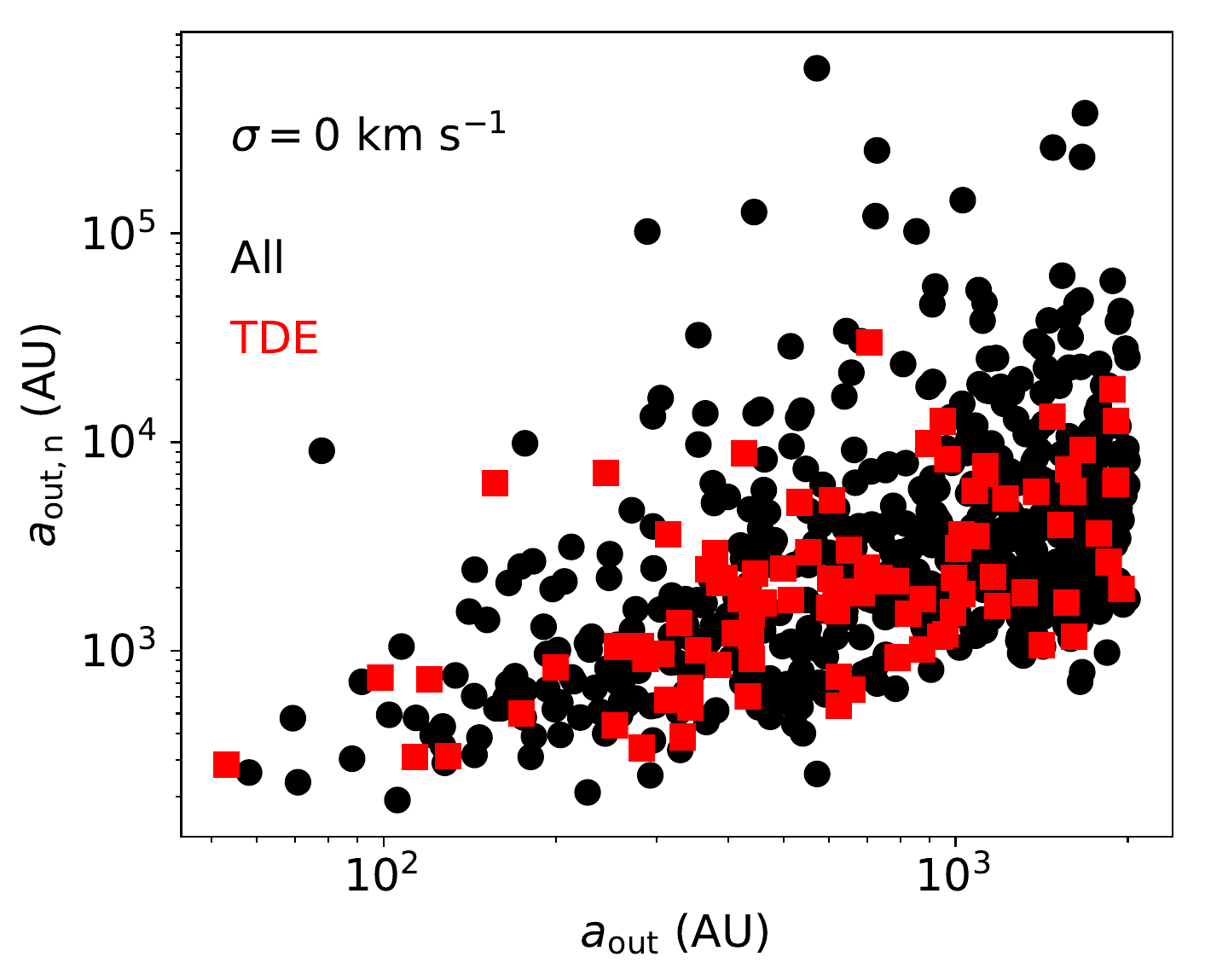}\\
\includegraphics[scale=0.55]{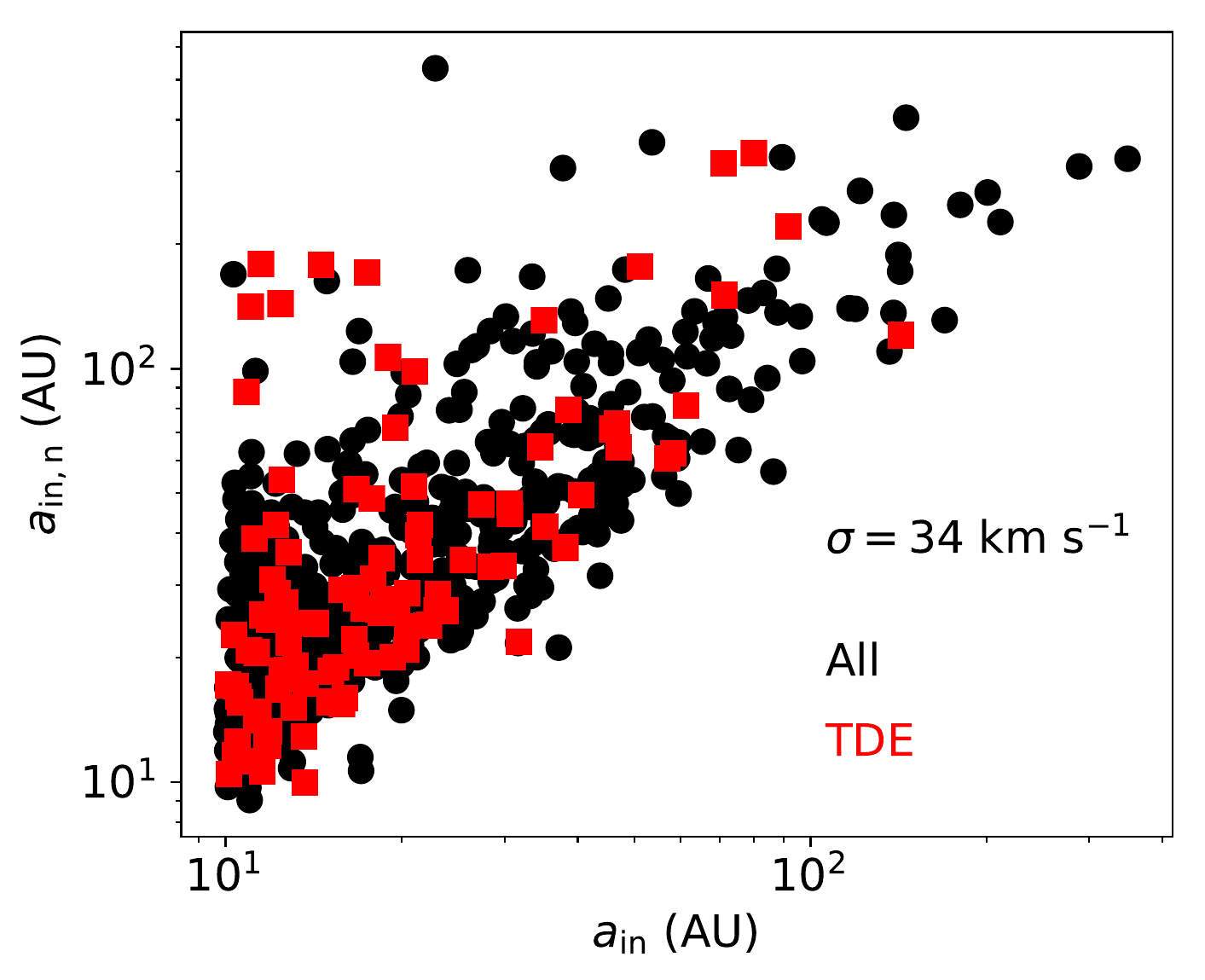}
\hspace{0.5cm}
\includegraphics[scale=0.55]{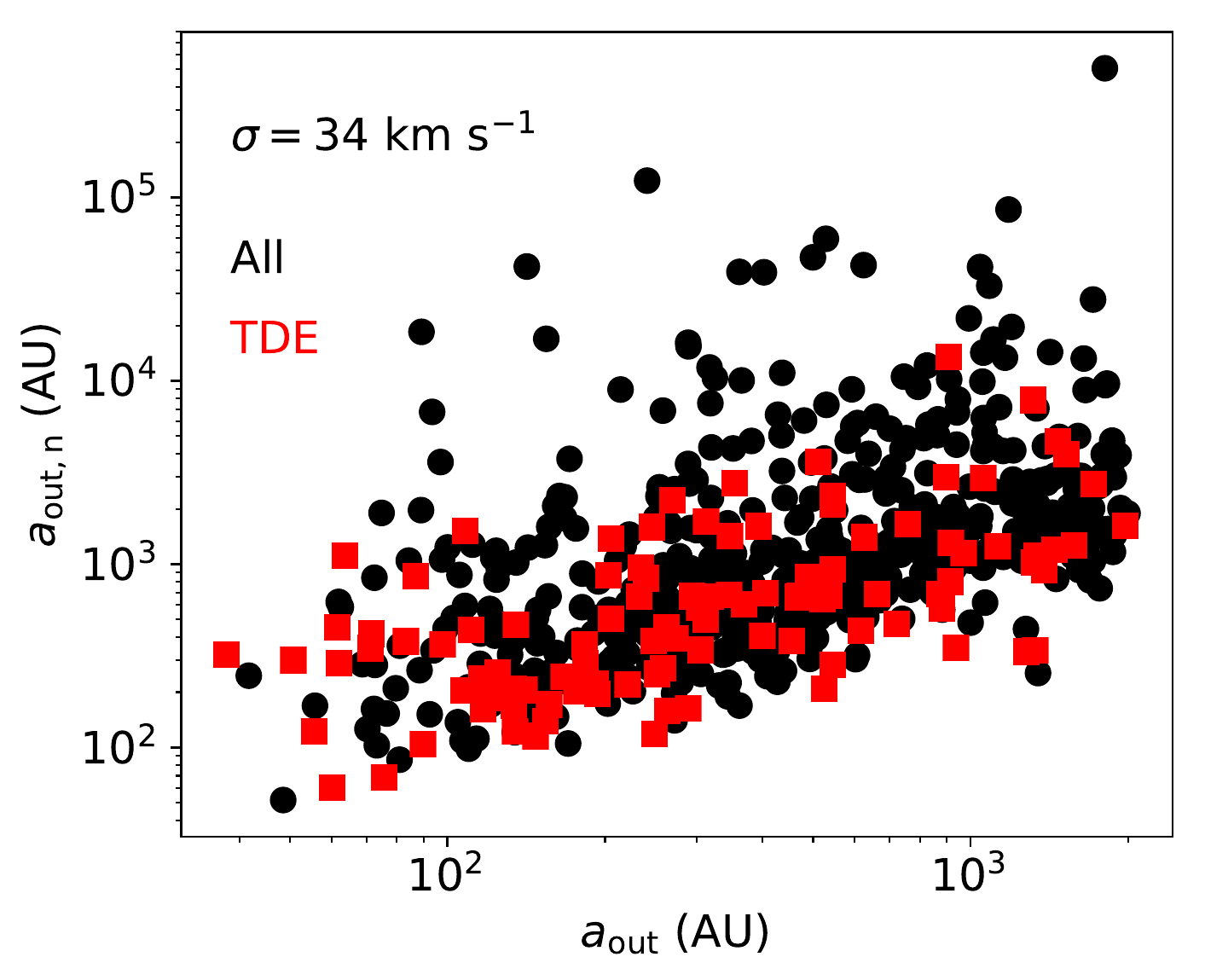}
\caption{Distributions of inner and outer semi-major axes of SBH-MS binaries in triples before ($\ain$ and $\aout$) and after ($a_{\rm in,n}$ and $a_{\rm out,n}$) the SN events for $\sigma=0\kms$ and $\sigma=34\kms$ kicks. Systems that lead to a TDE are marked in red.
}
\label{fig:ainoutsig}
\end{figure*}

We consider a hierarchical triple system that consists of an inner binary of mass $m_{\rm in}=m_1+m_2$ and a third body of mass $m_3$ that orbits the inner binary \citep[for details see][]{pijloo2012}. The triple can be described in terms of Keplerian orbital elements for both the inner orbit (i.e., the relative motion of $m_1$ and $m_2$), and the outer orbit (i.e., the relative motion of $m_3$ around the centre of mass of the inner binary). The semi-major axis and eccentricity of the inner orbit are $\ain$ and $\ein$, respectively, while the semi-major axis and eccentricity of the outer orbit are $\aout$ and $\eout$, respectively. The inner and outer orbital planes have initial mutual inclination $i_0$.

When a star undergoes an SN event, we assume that it takes place instantaneously, i.e. on a time-scale much shorter than the orbital period \citep{toonen2016,lu2019}. At the time of detonation, the star instantaneously loses mass. Under this assumption, the position of the body that undergoes an SN event is assumed not to change\footnote{We assume the SN shell has no impact on the companion stars.}. However, due to asymmetric mass loss, the exploding star is imparted a kick to its centre of mass \citep{bla1961}. We assume that the velocity kick is drawn from a Maxwellian distribution,
\begin{equation}
p(\vk)\propto \vk^2 e^{-\vk^2/\sigma^2}\ ,
\label{eqn:vkick}
\end{equation}
where $\sigma$ is the root-mean-square kick velocity which characterizes the distribution\footnote{Here, we do not take into account the fact that natal kicks for BHs can be reduced by the amount of fallback as $v_{\rm k,BH}=\vk(1-f_{\rm fb})$, where $f_{\rm fb}$ is the fallback parameter.}.

We assume that the SN event happens first in the primary star of the inner binary. Before the SN takes place, for the inner binary consisting of two stars with masses $m_1$ and $m_2$, a relative velocity $v=|{\bf{v}}|=|\bf{v}_1-\bf{v}_2|$, and separation distance $r=|{\bf{r}}|=|\bf{r}_1-\bf{r}_2|$, energy conservation  implies
\begin{equation}
|{\bf{v}}|^2=\mu\left(\frac{2}{r}-\frac{1}{\ain}\right)\ ,
\label{eqn:vcons}
\end{equation}
where $\mu=G(m_1+m_2)$. The angular momentum integral $\bf{h}$ is related to the orbital parameters by
\begin{equation}
|{\bf{h}}|^2=|{\bf{r}}\times {\bf{v}}|^2=\mu \ain(1-e_{\mathrm{in}}^2)\ .
\label{eqn:hcons}
\end{equation}

After the SN event, the orbital semi-major axis and eccentricity change due to the mass loss $\Delta m$ in the primary star and the natal kick $\vk$ (which is assumed to be isotropic). The total mass of the binary decreases to $m_{\rm in,n}=m_{1,n}+m_2$, where $m_{1,n}=m_1-\Delta m$. The relative velocity becomes ${\bf{v_n}}={\bf{v}}+{\bf{\vk}}$, while ${\bf{r_\mathrm{n}}}={\bf{r}}$. We assume
that the SN takes place instantaneously. The new semi-major axis can be computed from Eq.~\eqref{eqn:vcons},
\begin{equation}
a_{\rm in,n}=\left(\frac{2}{r}-\frac{v_n^2}{\mu_{\rm in,n}}\right)^{-1}\ ,
\end{equation}
where $\mu_{\rm in,n}=G(m_{1,n}+m_2)$, and the new eccentricity can be computed from Eq.~\eqref{eqn:hcons},
\begin{equation}
e_{\rm in,n}=\left(1-\frac{|{\bf{r}}\times {\bf{v_n}}|^2}{\mu_{\rm in,n} a_{\rm in,n}}\right)^{1/2}\ .
\end{equation}

Since the primary in the inner binary undergoes an SN event, an effective kick ${\bf{V_{\rm cm}}}$ is imparted to its centre of mass \citep{pijloo2012}. As a consequence of the mass loss in the primary, the centre of mass position of the inner binary
\begin{equation}
{\bf{r_{\rm cm}}}=\left(1-\frac{m_2}{m_{\rm in}}\right){\bf{r}_1}+\frac{m_2}{m_{\rm in}}{\bf{r}_2}
\end{equation}
changes due to an instantaneous translation to ${\bf{r_{\rm cm,n}}}={\bf{r_{\rm cm}}}+{\bf{\Delta r_{\rm cm}}}$, where
\begin{equation}
{\bf{\Delta r_{\rm cm}}}=\left(\frac{m_2}{m_{\rm in,n}}-\frac{m_2}{m_{\rm in}}\right){\bf{r}}\ .
\end{equation}
Thus, the separation distance ${\bf{R_{\rm 3}}}$ between the centre of mass of the inner binary and the tertiary star becomes
\begin{equation}
{\bf{R_{\rm 3,n}}}={\bf{R_{\rm 3}}}+{\bf{\Delta r_{\rm cm}}}\ ,
\end{equation}
while the relative velocity ${\bf{V_{\rm 3}}}$ changes into
\begin{equation}
{\bf{V_{\rm 3,n}}}={\bf{V_{\rm 3}}}+{\bf{V_{\rm cm}}}\ ,
\end{equation}
where
\begin{equation}
{\bf{V_{\rm cm}}}=\left(1-\frac{m_2}{m_{\rm in,n}}\right)\left({\bf{v}_1}+{\bf{v}_{\rm k}}\right)+\frac{m_2}{m_{\rm in,n}}{\bf{v}_2}\,.
\end{equation}
The outer semi-major axis $\aout$ and eccentricity $\eout$ change accordingly. To compute the relative change with respect to the pre-SN values, the strategy is to use again Eq.~\eqref{eqn:vcons} and Eq.~\eqref{eqn:hcons} for the outer orbit,
\begin{equation}
a_{\rm out,n}=\left(\frac{2}{R_{\rm 3,n}}-\frac{V_{\rm 3,n}^2}{\mu_{\rm out,n}}\right)^{-1}\ ,
\end{equation}
where $\mu_{\rm out,n}=G(m_{1,n}+m_2+m_3)$,
\begin{equation}
e_{\rm out,n}=\left(1-\frac{|{\bf{R_{\rm 3,n}}}\times {\bf{V_{\rm 3,n}}}|^2}{\mu_{\rm out,n} a_{\rm out,n}}\right)^{1/2}\ .
\end{equation}

Finally, the inclination of the outer binary orbital plane with respect to the inner binary orbital plane is tilted due to the kick. The new relative inclination $i_{\rm n}$ is computed from 
\begin{equation}
i_{\rm n}=\arccos \left(\frac{\bf{L_{\rm n}}\cdot \bf{L_{\rm 3,n}}}{L_{\rm n}\ L_{\rm 3,n}} \right) \ ,
\end{equation}
where $\bf{L_{n}}$ and $\bf{L_{\rm 3,n}}$ are the post-SN angular-momentum vectors of the inner and outer orbit, respectively.

In the event that either of the other two stars in the triple explode, or both of them, the same prescriptions described above can be applied to compute the post-SN orbital parameters. After every SN event, if either $\ain\le 0$ or $\aout\le 0$ the triple becomes unbound. 

In this paper we consider triple systems that survive the SN events and where the inner binary is made up of a SBH and a MS star of masses $\mbh$ and $\mms$, respectively. A triple system undergoes KL oscillations in eccentricity whenever the initial mutual orbital inclination of the inner and outer orbits is in the window $i\sim 40^\circ$-$140^\circ$ \citep{koz62,lid62}. At the quadrupole order of approximation, the KL oscillations occur on a timescale \citep{nao16},
\begin{equation}
T_{\rm KL}=\frac{8}{15\pi}\frac{m_{\rm tot}}{m_{\rm 3}}\frac{P_{\rm out,n}^2}{P_{\rm SBHMS}}\left(1-e_{\rm out,n}^2\right)^{3/2}\ ,
\end{equation}
where $m_{\rm 3}$ is the mass of the outer body orbiting the inner SBH-MS binary, $m_{\rm tot}$ is the total mass of the triple system, and $P_{{\rm SBHMS}}\propto a_{{\rm in,n}}^{3/2}$ and $P_{{\rm out,n}}\propto a_{{\rm out,n}}^{3/2}$ are the orbital periods of the inner SBH-MS binary and of the outer binary, respectively. In the quadrupole interaction approximation, the maximal eccentricity is a function of the initial mutual inclination,
\begin{equation}
e_{\rm in,n}^{\max}=\sqrt{1-\frac{5}{3}\cos^2 i_\mathrm{n}}\ .
\label{eqn:emax}
\end{equation}
Whenever $i_{\rm n}\sim 90^\circ$, the inner binary eccentricity approaches almost unity. 
If the octupole corrections are taken into account and the outer orbit is eccentric, the inner eccentricity can reach almost unity even if the initial inclination is outside of the $i_{\rm n}\sim 40^\circ$-$140^\circ$ KL range \citep{naoz13a,li14}. However, KL oscillations can be suppressed by additional sources of precession, such as tidal bulges or relativistic precession \citep{naoz13b,nao16}.

We note that these analytical relations are useful to give a qualitative understanding of the evolution of triple systems, but we are not limited to these secular prescriptions since we use $N$-body simulations, as discussed in the next section. Moreover, the system may become non-hierarchical in some cases and the above secular equations lose validity \citep[e.g.][]{antognini14,fragrish2018}.

If the inner SBH-MS binary reaches sufficiently high eccentricity, the star can be tidally disrupted by the SBH. This occurs whenever their relative distance is smaller than the tidal disruption radius,
\begin{equation}
R_T=R_* \left(\frac{\mbh}{\mms}\right)^{1/3}\ ,
\label{eqn:rtid}
\end{equation}
where $R_*$ is the radius of the star that we compute from \citet{dem91},
\begin{equation}
R_*=
\begin{cases}
1.06\ (\mms/\msun)^{0.945}\rsun& \text{$ \mms< 1.66\msun$}\ ,\\
1.33\ (\mms/\msun)^{0.555}\rsun& \text{$ \mms> 1.66\msun$}\ .
\end{cases}
\end{equation}
We note that the exact value of Eq,~\ref{eqn:rtid} could depend on the properties of the star \citep{Guillochon2013}. Moreover, the effective tidal radius could be substantially larger owing to the fact that the stellar spin would increase at any close passages to the SBH \citep{goli2019}, if the merger happens after several KL cycles.

\section{N-body simulations}
\label{sect:results}

\begin{table}
\caption{Models: name, mean of SBH kick-velocity distribution ($\sigma$), eccentricity distribution ($f(e)$), maximum outer semi-major axis of the triple ($\amax$), fraction of TDEs from the $N$-body simulations ($f_{\rm TDE}$).}
\centering
\begin{tabular}{lccccc}
\hline
Name & $\sigma$ ($\kms$) & $f(e)$ & $\amax$ (AU) & $f_{\rm TDE}$\\
\hline\hline
A1 & $34$ & uniform  & $2000$ & $0.14$\\
A2 & $0$  & uniform  & $2000$ & $0.14$\\
A3 & $13$ & uniform  & $2000$ & $0.14$\\
B1 & $34$ & uniform  & $5000$ & $0.12$\\
B2 & $34$ & uniform  & $7000$ & $0.13$\\
C1 & $34$ & thermal  & $2000$ & $0.11$\\
\hline
\end{tabular}
\label{tab:models}
\end{table}

\subsection{Initial conditions}

The stellar triples in our simulations are initialized as follows. In total, we consider six different sets of initial conditions (see Table~\ref{tab:models}).

For simplicity, we assume that stars in the mass range $20 \msun$--$150\msun$ collapse to a SBH. In all our models, we adopt the \citet{kroupa2001} initial mass function
\begin{equation}
f(m) = 0.0795\,\msun^{-1} (m/\msun)^{-2.3}\quad{\rm if~} m \geq 0.5\msun
\label{eqn:bhmassfunc}
\end{equation}
in the relevant mass range\footnote{The constant coefficient accounts for the fraction of stars with $m< 0.5\msun$ such that the integral of $\int_0^{\infty} f(m) dm = 1$.}. We sample the mass $m_1$ of the most massive star in the inner binary in the mass range $20\msun$-$150\msun$, reflecting the progenitor of the SBH. We assume that the primary produces a SBH of mass $\mbh=m_1/3$. We adopt a flat mass ratio distribution for both the inner orbit, $q_{12}=m_2/m_1$, and the outer orbit, $q_{123}=m_3/(m_1+m_2)$ \citep{sana12,duch2013,sana2017}\footnote{From observations, \citet{duch2013} found that $f(q)\propto q^{1.16\pm0.16}$ and $q^{-0.01\pm0.03}$ for solar type stars with period less than or larger than $10^{5.5}\,$day, respectively, while \citet{sana13} found $f(q)\propto q^{-1.0\pm 0.4}$ for massive O-type stars.}. We sample $q_{12}$ in the range $(0.5\msun/m_1)$-$(8\msun/m_1)$ so that the mass of the secondary ($\mms=q_{12} m_1$) in the inner binary is in the range $0.5\msun$-$8\msun$, and sample $q_{123}$ in the range $(0.5\msun/(m_1+m_2))$-$(150\msun/(m_1+m_2))$ so that the mass of the third companion ($m_3=q_{123}(m_1+m_2)$) is in the range $0.5\msun$-$150\msun$. If the initial mass of the tertiary $m_3$ is in the range $8 \msun$--$20\msun$, we assume it will form a NS of mass $1.3\msun$. For higher masses, we assume that it collapses to a SBH of mass $m_3/3$. For comparison, we also estimate how the final TDE rate in triples changes if the mass ratio distribution is assumed to be log-uniform \citep{sana13}.

The exact values of the NS and SBH masses vary with the specific evolutionary path (i.e. single, binary) followed by the progenitor star, and  depend on a number of variables, such as metallicity, stellar winds, rotation, and possible common envelope phases. For example, stellar winds could be important since they may change the orbital parameters of binaries before a SBH or a NS is formed, as well as contribute to the mass loss of a star during the main sequence;  their strength depends on the progenitor metallicity, and hence these variables all contribute to determine the final mass of the NSs and SBHs. Here we are not modeling these effects, which could reduce the available parameter space \citep{shapp2013}. The previous picture can become even more complicated if mass loss during possible episodes of Roche-lobe overflows and common evolution phases in the triple are taken into account. While these processes are relatively well understood in binaries, they are not modeled in a self-consistent way in triple systems because of the possible interplay with the KL cycles \citep{rosa2019,hamd2019}. Therefore, we prefer to adopt a simple procedure to derive the final SBH and NS masses,   where the maximum SBH mass is consistent with recent theoretical results on pulsational pair instabilities that limit the maximum BH mass to $\sim 50\msun$ \cite{bel2016b}. 

The distributions of the inner and outer semi-major axes, $\ain$ and $\aout$ (respectively), are assumed to be flat in log-space (\"{O}pik's law), consistent with the results of \citet{kob2014}. We set as a minimum separation $10$ AU, and adopt different values for the maximum separation $\amax=2000$ AU--$5000$ AU--$7000$ AU \citep{sana2014}. For the orbital eccentricities of the inner and outer binaries, or $\ein$ and $\eout$ (respectively), we assume flat distributions. For comparison, we run one additional model where we consider a thermal eccentricity distribution.

The initial mutual inclination $i_0$ between the inner and outer orbits is drawn from an isotropic distribution (i.e. uniform in $\cos i_0$). The other relevant angles are drawn from uniform distributions.

After sampling the relevant parameters, we check that the initial configuration satisfies the stability criterion for stable hierarchical triples \citep{mar01},
\begin{equation}
\frac{R_{\rm p}}{a_{\rm in}}\geq 2.8 \left[\left(1+\frac{m_{\rm 3}}{m_1+m_2}\right)\frac{1+\eout}{\sqrt{1-\eout}} \right]^{2/5}\left(1.0-0.3\frac{i_0}{\pi}\right)\ ,
\label{eqn:stabts}
\end{equation}
where $R_p=\aout(1-\eout)$ is the pericentre distance of the outer orbit.

If the systems are stable according to the previous criterion, we let the primary star in the inner binary undergo an SN explosion and instantaneously convert it to a SBH. The orbital elements of the inner and outer orbit are updated following the procedure discussed in Section~\ref{sect:supern}, to account both for mass loss and the natal kick.

The distribution of natal kick velocities of SBHs and NSs is unknown. To be conservative, we implement momentum-conserving kicks. As a consequence, the mean kick velocities for SBHs are lower relative to those of NSs by a factor $\mbh/\mns$. In our fiducial model, we consider a non-zero natal kick velocity for the newly formed SBHs, by adopting Eq.~\eqref{eqn:vkick} with $\sigma=34 \kms$ for $\mbh=10\msun$. This corresponds to a mean velocity $\sim 34 \kms \times \mbh/\mns \approx 260 \kms$ for NSs, consistent with the distribution deduced by \citet{hobbs2005}. We run an additional model where we adopt $\sigma=13 \kms$, which translates into a mean velocity $\sim 13 \km \times \mbh/\mns \approx 100 \kms$ for NSs, consistent with the distribution of natal kicks found by \citet{arz2002}. Finally, we adopt a model where no natal kick is imparted during SBH formation (which we label $\sigma=0 \kms$). We note that even in this case, the triple experiences a kick to its centre of mass at the time of SBH formation, because one of the massive components suddenly loses mass \citep{bla1961}.

If the third companion is more massive than $8\msun$, we let it undergo an SN event and convert to a compact object, either a NS with $m_{\rm 3,n}=1.3\msun$ or a SBH with $m_{\rm 3,n}=m_3/3$\footnote{We do not model the process that leads to the formation of a white dwarf. We note that in the case the third companion becomes a white dwarf, and the system remains bound according to Eq.~\eqref{eqn:stabts}, some of the systems could still produce a TDE.}. If the triple remains bound, we check again the triple stability criterion of \citet{mar01} with the updated masses and orbital parameters for the inner and outer orbits. Figure~\ref{fig:ainoutsig} shows the distributions of inner and outer semi-major axes of stable SBH-MS binaries in triples before ($\ain$ and $\aout$) and after ($a_{\rm in,n}$ and $a_{\rm out,n}$) the SN events, for $\sigma=0\kms$ and $\sigma=34\kms$.

Given the above set of initial parameters, we integrate the equations of motion of the 3-bodies,
\begin{equation}
{\ddot{\textbf{r}}}_i=-G\sum\limits_{j\ne i}\frac{m_j(\textbf{r}_i-\textbf{r}_j)}{\left|\textbf{r}_i-\textbf{r}_j\right|^3}\ ,
\end{equation}
with $i=1$,$2$,$3$, by means of the \textsc{ARCHAIN} code \citep{mik06,mik08}, a fully regularized code able to model the evolution of binaries of arbitrary mass ratios and eccentricities with high accuracy, thanks to a transformed leapfrog method combined with the Bulirsch-Stoer extrapolation. \textsc{ARCHAIN} includes PN corrections up to order PN2.5 and a treatment of equilibrium dissipative and non-dissipative tides \citep{hut1981A&A....99..126H,fan2019}. For MS stars, we fix the apsidal motion constant to $k=0.014$ and the time-lag factor to $\tau_{\rm lag}=0.66$ s \citep{fan2019}. We perform $1000$ simulations for each model, for the initial conditions provided in Table~\ref{tab:models}. If the MS star is tidally disrupted (see Eq.~\eqref{eqn:rtid}), we stop the integration. Otherwise, for the secondary star in the inner binary $m_2$ and the third companion $m_3$ (assuming it does not evolve to form a compact object, or $m_3< 8\msun$), we compute a MS lifetime. This is simply parameterized as \citep[e.g.][]{iben91,hurley00,maeder09},
\begin{equation}
\tau_{\rm MS} = \max(10\ (m/\msun)^{-2.5}\,{\rm Gyr}, 7\,{\rm Myr})\ .
\end{equation}
We fix the maximum integration time,
\begin{equation}
T=\left[\min \left(\tau_{\rm MS,1}, \tau_{\rm MS,3} \right),\min \left(10^3 \times T_{\rm KL}, 10\ \gyr \right)\right]\,,
\label{eqn:tint}
\end{equation}
where $T_{\rm KL}$ is the triple KL timescale. We also check if the secondary star in the inner binary or the third star (if it does not collapse to a compact object) overflow their Roche lobe using the formulae provided in \citet{egg83}. In this case, we stop the integration.

\subsection{Results}

\begin{figure} 
\centering
\includegraphics[scale=0.55]{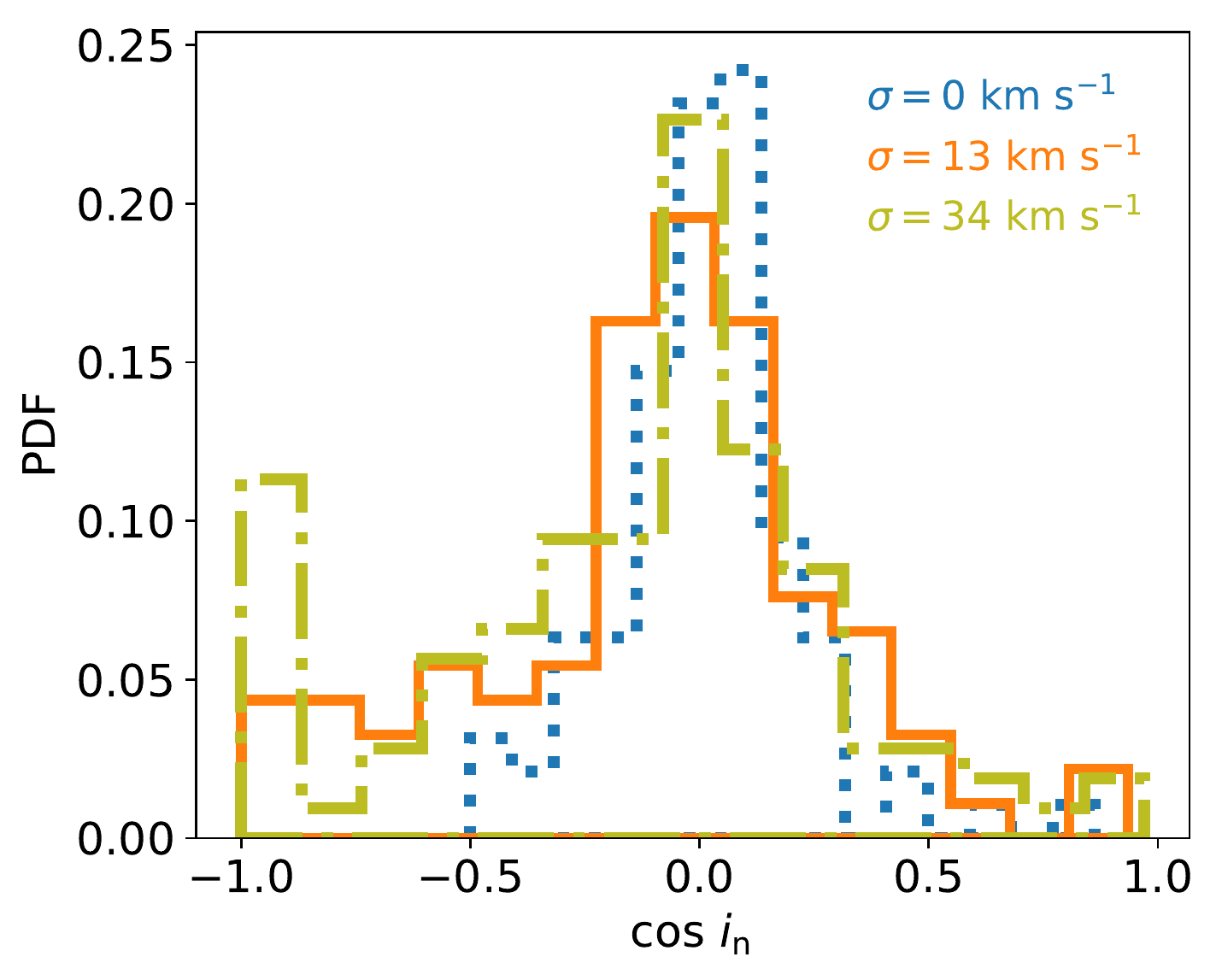}
\caption{Inclination PDF of merging SBH-MS binaries in triples, for $\amax=2000$ AU and different values of the mean kick velocity $\sigma$.}
\label{fig:incl}
\end{figure}

A SBH-MS binary is expected to be significantly perturbed by the tidal field of the third companion whenever its orbital plane is sufficiently inclined with respect to the outer orbit \citep{lid62,koz62}. According to Eq.~\eqref{eqn:emax}, the SBH-MS eccentricity reaches almost unity when $i_{\rm n}\sim 90^\circ$. Figure~\ref{fig:incl} shows the inclination probability distribution function (PDF) of merging SBH-MS binaries in triples. The distributions are shown for $\amax=2000$ AU and different values of $\sigma$. Independently of the mean of the natal kick velocity, the majority of SBH-MS TDEs in triples occur when the inclination approaches $\sim 90^\circ$. In this case, the KL effect is maximal, leading to eccentricity oscillations up to unity. SBH-MS systems that undergo a TDE with low relative inclinations typically have large initial eccentricities. We also identify a possible second peak at $i_n \sim 180^{\circ}$ for high natal kicks, which corresponds to a coplanar counterrotating configuration. We note that an enhancement of GW driven mergers in SBH triples at $\sim 180^{\circ}$ was also seen previously in non-hierarchical three-body simulations by \citet{arc2018} which may be related to the coplanar flip phenomenon \citep{li14}.

\begin{figure} 
\centering
\includegraphics[scale=0.55]{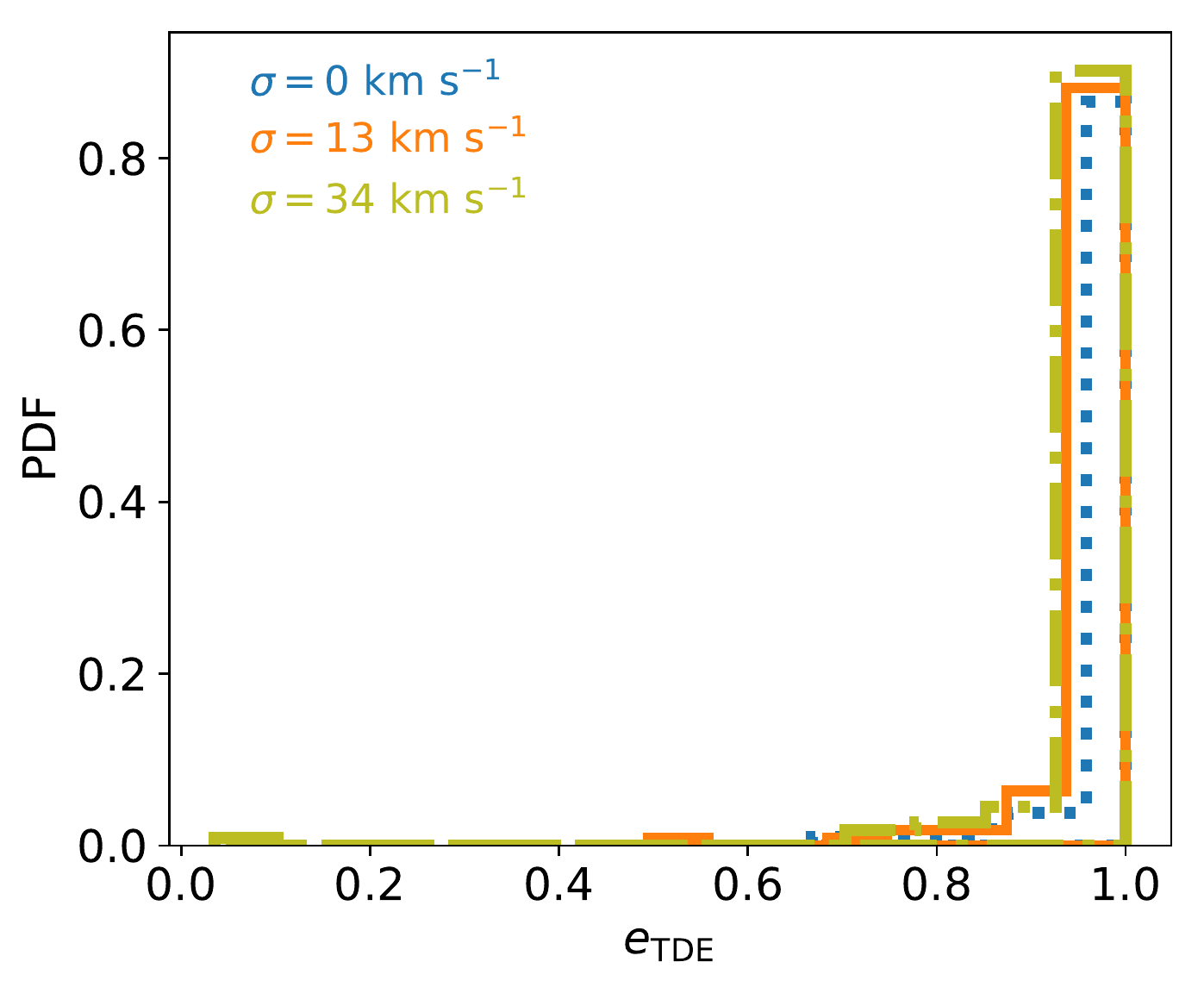}
\includegraphics[scale=0.55]{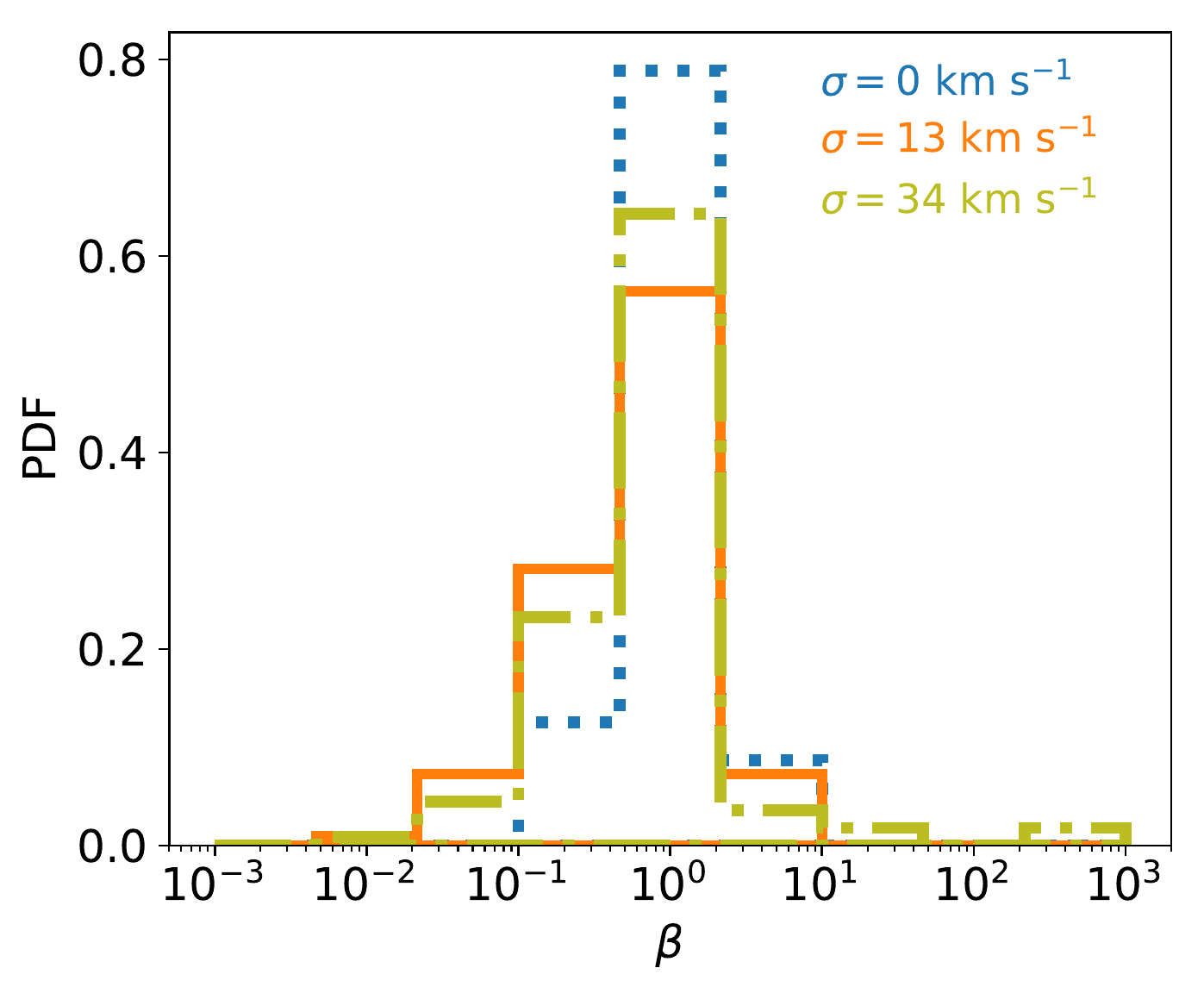}
\caption{PDF distribution of the eccentricity (top) and penetration factor (bottom) of MS stars that produce a TDE in triples, for $\amax=2000$ AU and different values of the mean kick velocity $\sigma$.}
\label{fig:betaeted}
\end{figure}

In Figure~\ref{fig:betaeted} we show the PDF distribution of the eccentricity $e_{\rm TDE}$ (top) and penetration factor $\beta$ (bottom) of MS stars that produce a TDE in triples, for $\amax=2000$ AU and different values of the mean kick velocity $\sigma$. The penetration factor
\begin{equation}
\beta \equiv \frac{R_{\rm T}}{R_{\rm p}} = \left(\frac{R_{\rm p}}{R_{\rm *}}\right)^{-1}\left(\frac{\mbh}{\mms}\right)^{1/3}\ ,
\end{equation}
where $R_{\rm p}$ is the pericentre of the MS star, gives an indication of how extreme some of these encounters could be. For all the $\sigma$'s, the distribution of $e_{\rm TDE}$ is peaked at $\sim 1$, similarly to the case of a TDE by a SMBH. The distribution of penetration factors is peaked at $\sim 1$, with non-negligible tails in the range $\sim 0.02$--$10$. This is different than the case of SMBHs and IMBHs, for which $\beta \gg 1$ and strong tidal compression is possible.

\begin{figure} 
\centering
\includegraphics[scale=0.55]{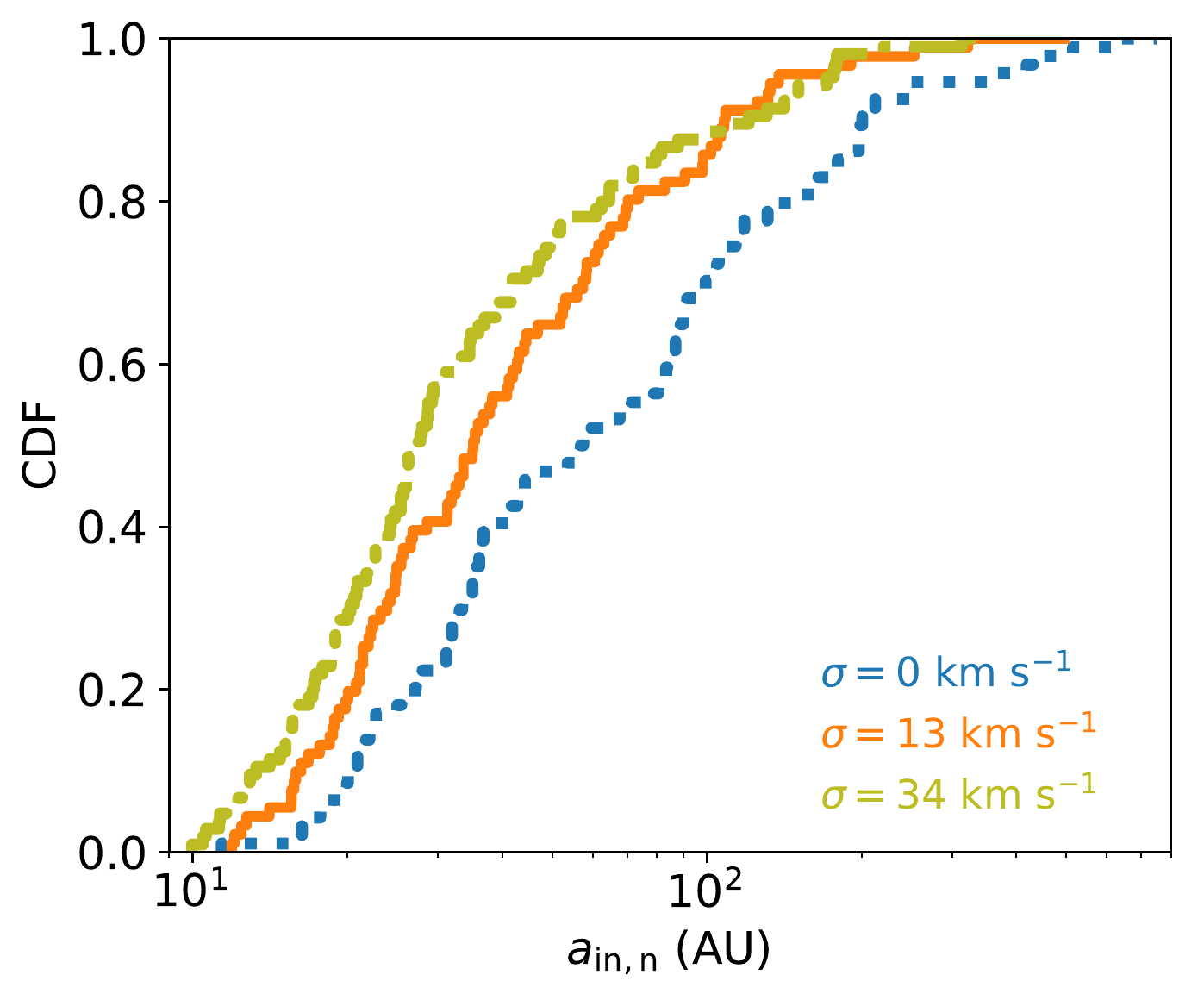}
\includegraphics[scale=0.55]{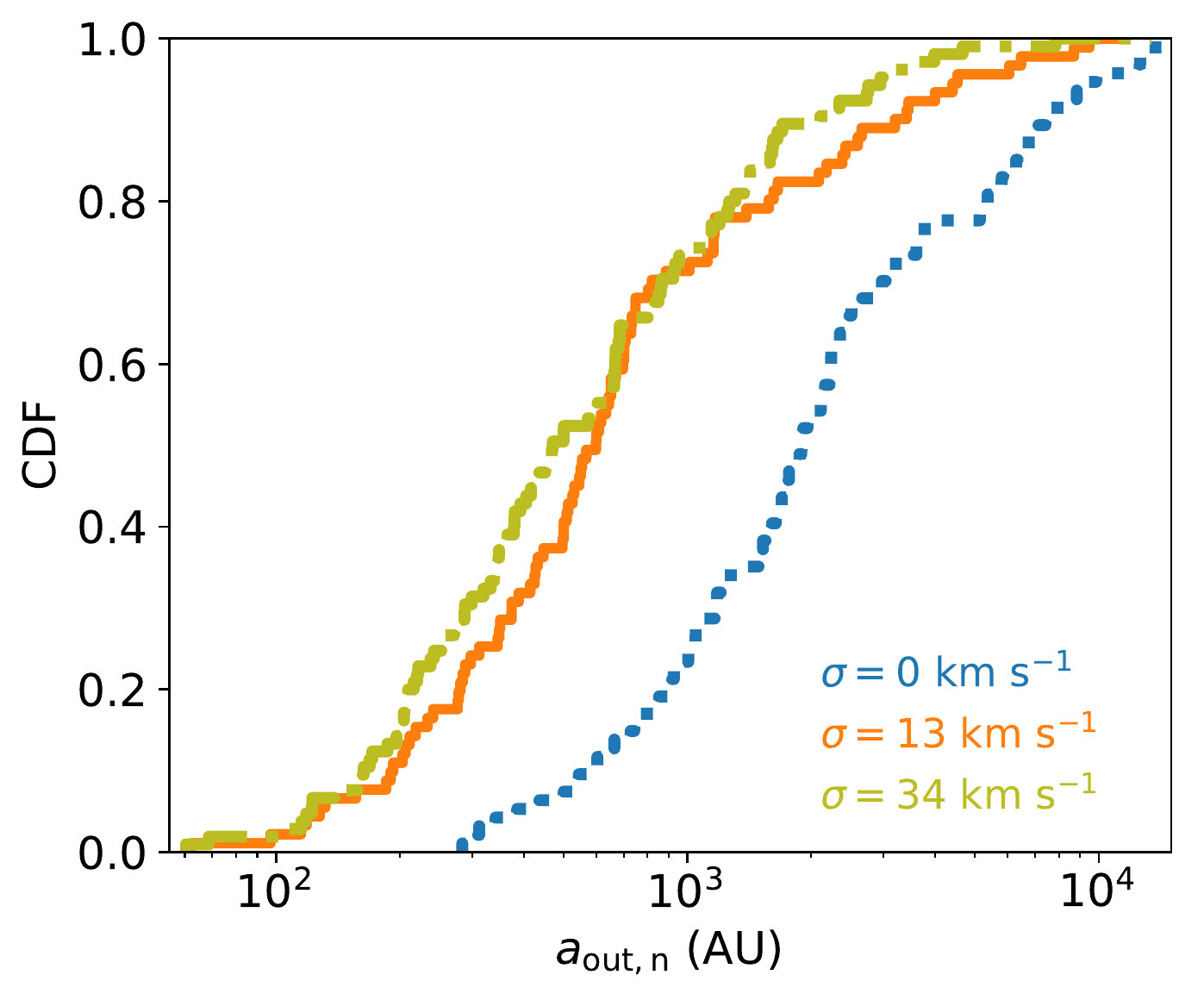}
\caption{Cumulative distribution function of inner (top) and outer (bottom) semi-major axes of SBH-MS binaries in triples that lead to a TDE, for different values of $\sigma$.}
\label{fig:ainaout}
\end{figure}

Figure~\ref{fig:ainaout} reports the cumulative distribution function (CDF) of the inner (top) and outer (bottom) semi-major axes of SBH-MS binaries in triples that lead to a TDE, for different values of $\sigma$. The larger the mean natal kick, the smaller the typical inner and outer semi-major axes. This can be understood by considering that triples with wide orbits are generally left unbound by large kick velocities, while they stay bound if the natal kick is not too intense. We find similar CDFs for $\sigma=13 \kms$ and $\sigma=34 \kms$ but significant differences for $\sigma=0 \kms$. For the inner orbit, we find that $\sim 50$\% of the systems that produce a TDE have $a_{\rm in,n}\lesssim 60$ AU and $\lesssim 30$ AU for $\sigma=0\kms$ and $34\kms$, respectively.  For the outer orbit, we find that $\sim 50$\% of the systems have $a_{\rm out,n}\lesssim 2000$ AU and $\lesssim 500$ AU for $\sigma=0\kms$ and $34\kms$, respectively.

\begin{figure} 
\centering
\includegraphics[scale=0.55]{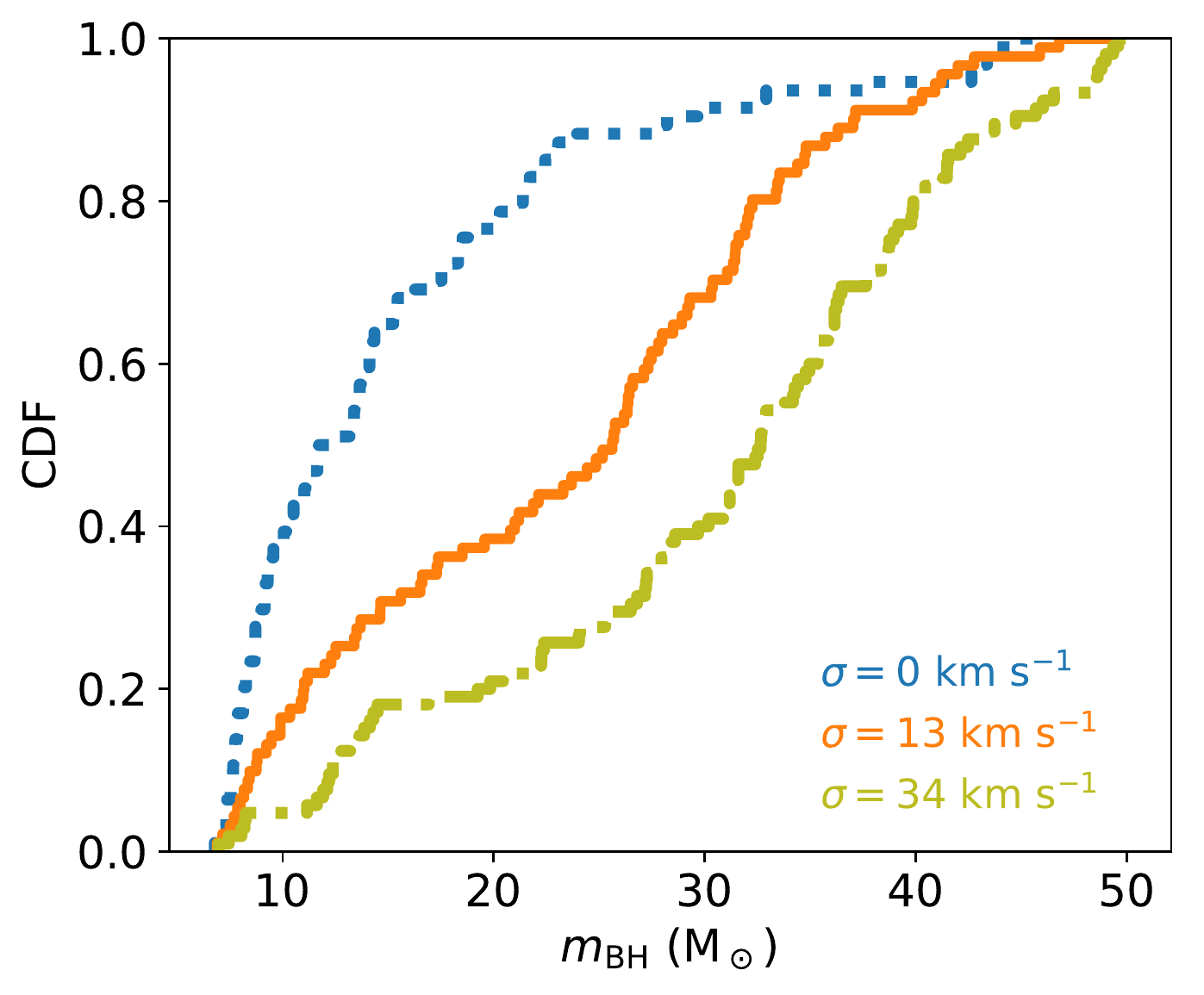}
\includegraphics[scale=0.55]{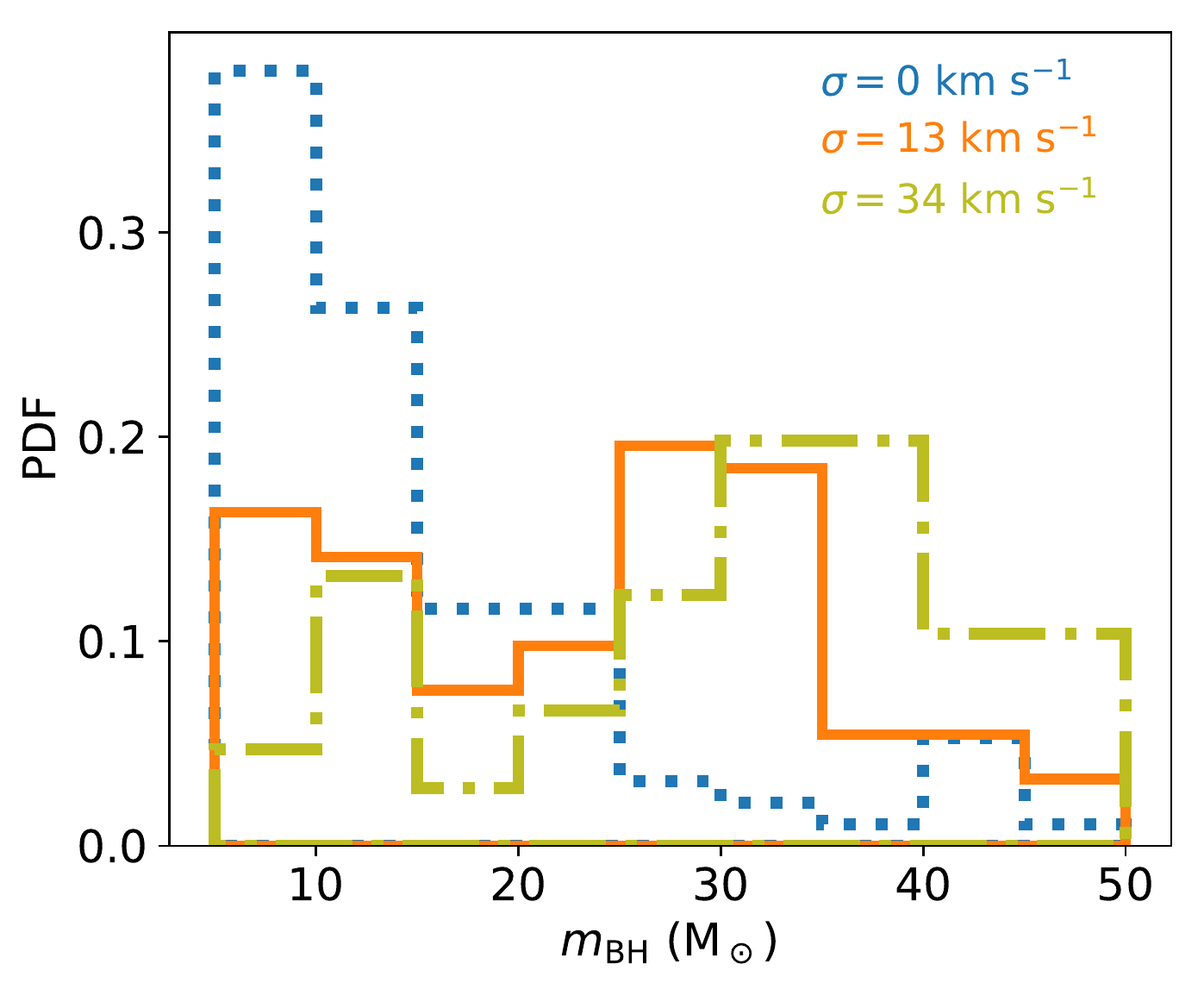}
\caption{CDF (top) and PDF (bottom) of the mass of SBHs in SBH-MS binaries in triples that lead to a TDE, for different values of $\sigma$.}
\label{fig:mbh}
\end{figure}

The typical mean natal kick velocity affects also the distribution of SBH masses in SBH-MS binaries that lead to a TDE in triple systems. We illustrate this in Figure~\ref{fig:mbh}, where we plot the CDF and PDF of $\mbh$ of SBH-MS binaries in triples that lead to a TDE, for different values of $\sigma$. In the case of $\sigma=0\kms$, we find that merging SBHs have typically lower masses compared to the models with $\sigma=13\kms$ and $\sigma=34\kms$. In the former case, $\sim 50\%$ of the SBHs that produce a TDE have masses $\lesssim 10\msun$, while for non-zero kick velocities we find that $\sim 50\%$ of the SBHs have masses $\lesssim 25\msun$ and $\lesssim 35\msun$ for $\sigma=13\kms$ and $\sigma=34\kms$, respectively. This is justified by our assumption of momentum-conserving kicks, where higher mass SBHs receive, on average, lower velocity kicks and, as a consequence, are more likely to be retained in triples and eventually produce a TDE.

\begin{figure} 
\centering
\includegraphics[scale=0.55]{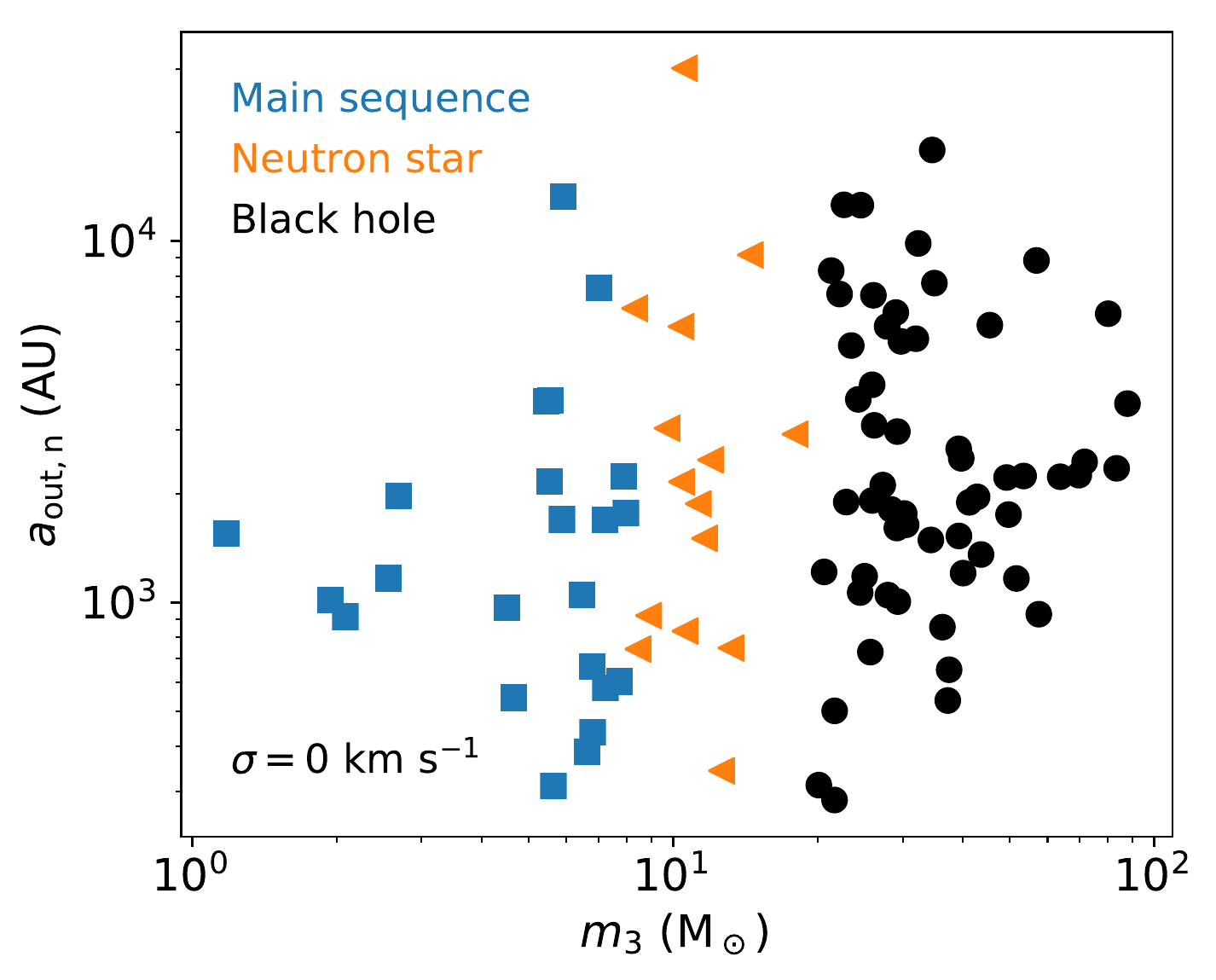}
\includegraphics[scale=0.55]{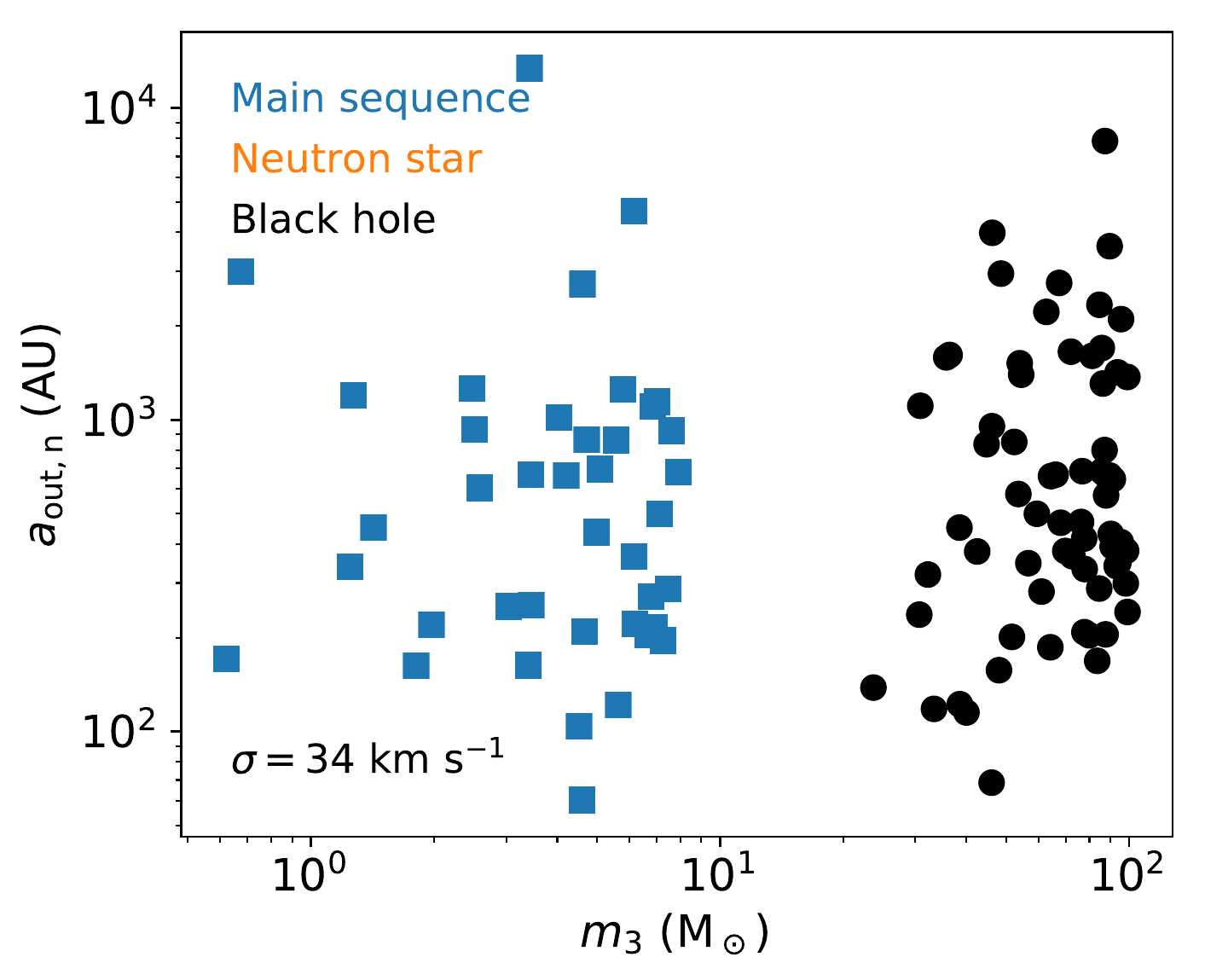}
\caption{Progenitor mass of the third companion of SBH-MS binaries in triples that lead to a TDE for $\sigma=0\kms$ (top) and $\sigma=34\kms$ (bottom).}
\label{fig:m3}
\end{figure}

The mean velocity $\sigma$ has an effect also in determining the typical mass of the third companion in triples that lead to a TDE. We show this in Figure~\ref{fig:m3}, where we plot the progenitor mass $m_3$ of the third companion of SBH-MS binaries in triples that lead to a TDE for $\sigma=0\kms$ (top) and $\sigma=34\kms$ (bottom). Note that $m_{\rm 3,n}=m_3$, if $m_3\lesssim 8\msun$. In the case of no natal kick velocity, the third companion in triples that lead to a TDE can be either a MS star, a NS, or a SBH. If $\sigma=34\kms$, which corresponds to a mean velocity of $\sim 260\kms$ for NSs, none of the third companions is a NS because the large $\vk$ typically unbinds the triple. Moreover, if the third companion is a SBH (i.e. given two SN events that lead to the formation of an SBH in the inner binary and an SBH as tertiary), its typical mass $m_{\rm 3,n}$ is smaller in the case $\sigma=0\kms$ than the case $\sigma=34\kms$, as explained above (for the SBH in the inner binary).

\begin{figure*} 
\centering
\includegraphics[scale=0.55]{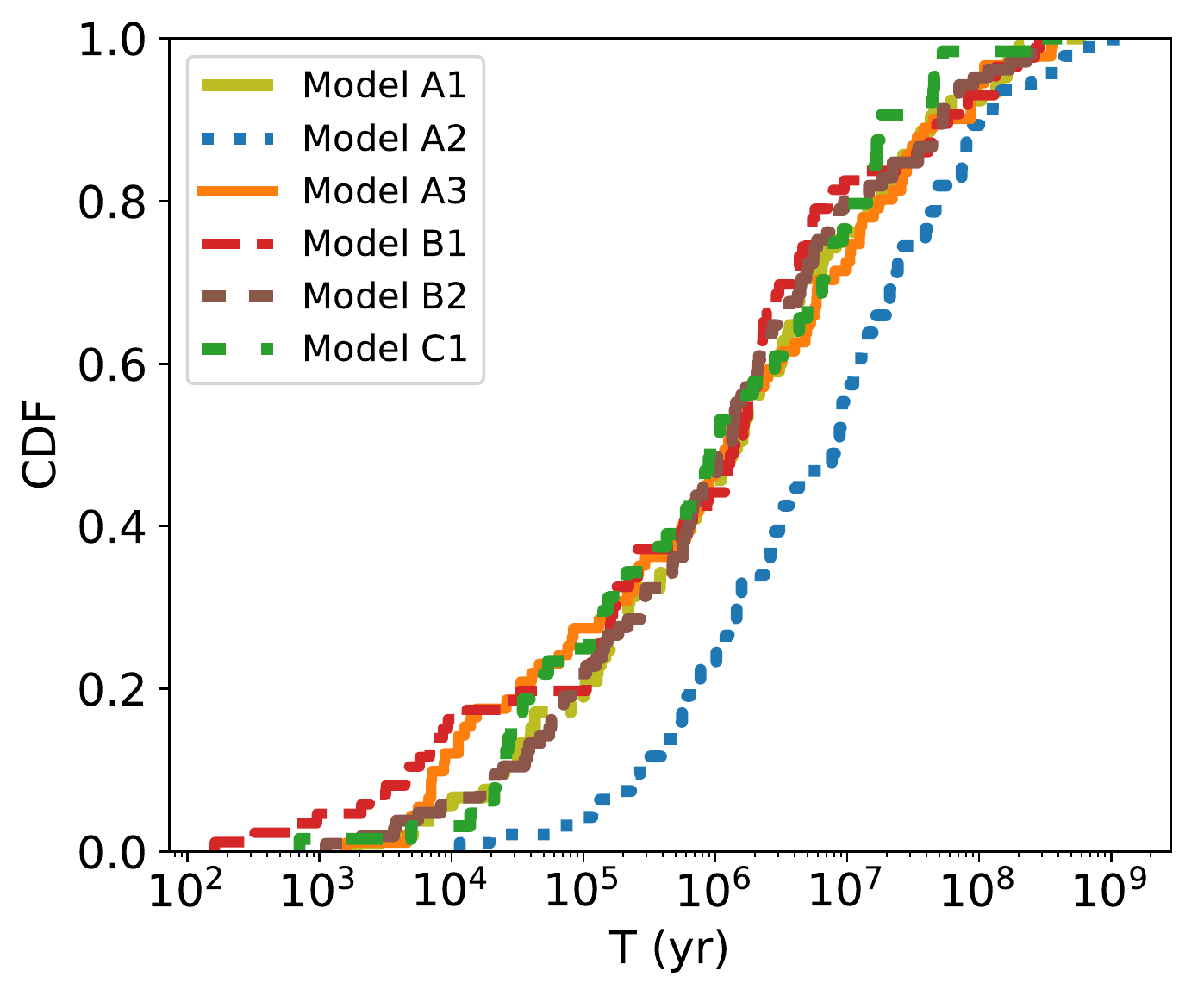}
\hspace{1cm}
\includegraphics[scale=0.55]{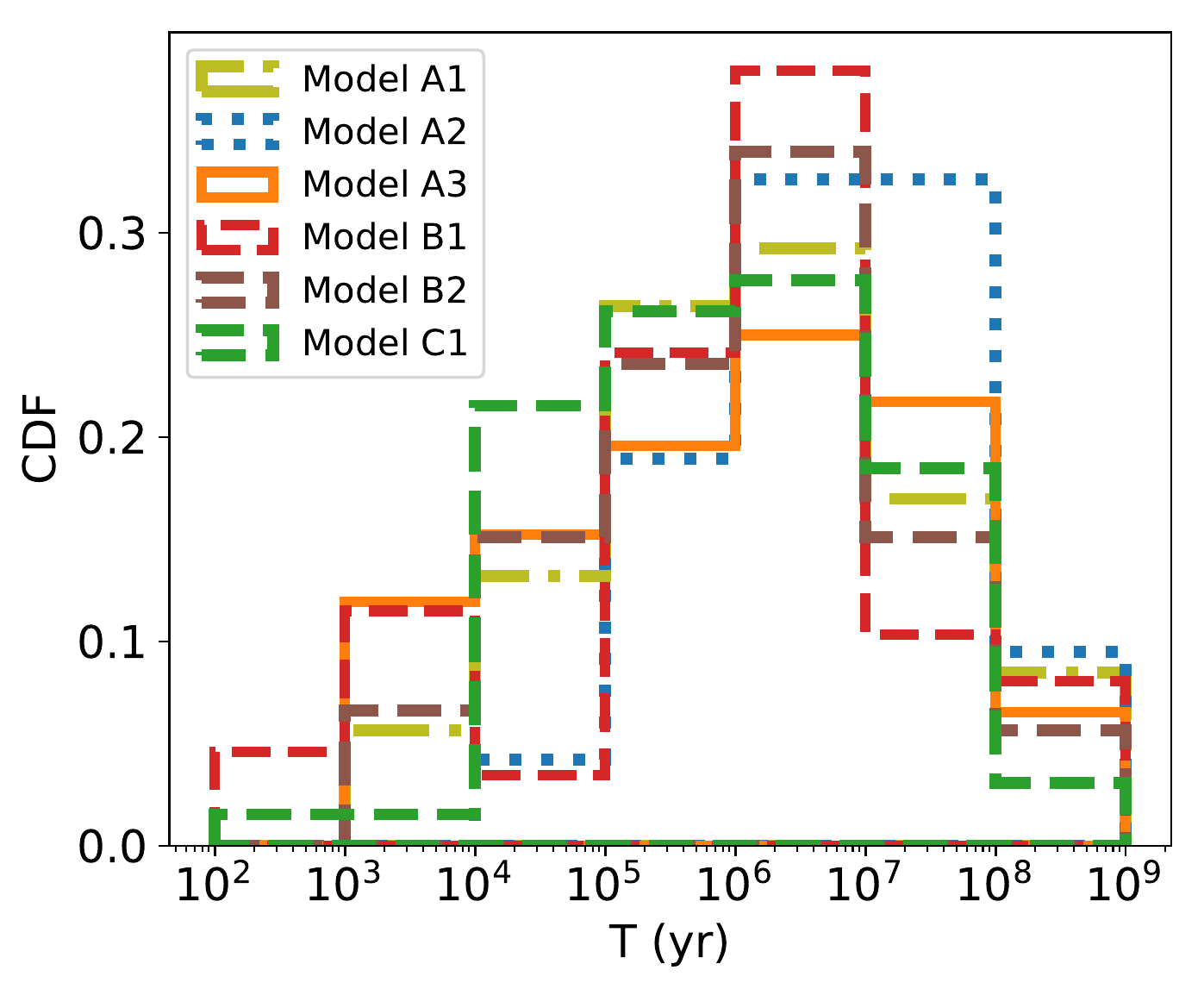}
\caption{CDF (left) and PDF (right) of the TDE time (after the SN event) of SBH-MS binaries in triples that lead to a TDE for all models (see Tab.\ref{tab:models}).}
\label{fig:ttde}
\end{figure*}

Figure~\ref{fig:ttde} reports the CDF and PDF of the TDE time (after the SN event) of SBH-MS binaries in triples that lead to a TDE for all models (see Tab.~\ref{tab:models}). The shape of these CDFs depends only on the value of $\sigma$, where Model A2 has a distribution shifted by about one order of magnitude compared to other models with non-zero natal kicks. In order to compute the TDE rate of SBH-MS in triples, we assume that the local star formation rate is $0.025 \msun$ Mpc$^{-3}$ yr$^{-1}$ \citep{both2011}, thus the number of stars formed per unit mass, volume, and time is given by,
\begin{align}
\dot{n}(m) &= \frac{\eta_{\rm SFR}\, f(m)}{\langle m\rangle} =
\nonumber\\&=
5.2\times 10^6 \left(\frac{m}{\msun}\right)^{-2.3}\ \mathrm{M}_\odot^{-1}\ \mathrm{Gpc}^{-3}\mathrm{yr}^{-1}\ ,
\end{align}
where $\langle m\rangle = 0.38 \msun$ is the average stellar mass. Assuming a constant star-formation rate, the TDE rate in triple systems is then,
\begin{align}
\mathcal{R}_\mathrm{TDE} &=
\eta (1-\zeta) f_{\rm 3} f_{\rm stable} f_{\rm TDE}\int_{20\msun}^{150\msun} \dot{n}(m_1) dm_1=
\nonumber\\&=
7.4\times 10^4 \eta (1-\zeta) f_{\rm 3} f_{\rm stable} f_{\rm TDE}\ \mathrm{Gpc}^{-3}\ \mathrm{yr}^{-1}\ ,
\end{align}
where $f_{\rm 3}$ is the fraction of stars in triples, $f_{\rm stable}$ is the fraction of systems that remain stable after the SN events take place, and $f_{\rm TDE}$ is the fraction of systems that lead to a TDE (see Tab.~\ref{tab:models}). The factor $\eta$ comes from assumptions that the secondary mass satisfies $1\msun \le m_2=q_{12}m_1\le 8\msun$, when sampling the mass ratio $q_{12}$,
\begin{equation}
    \eta = \frac{ \int_{0.01\msun}^{150\msun} d m_1 f_{\rm IMF}(m_1) \int_{{1\msun}/{m_1}}^{{{8\msun}/{m_1}}} dq  f_{q}(q) 
    }{
     \int_{20 \msun}^{150\msun} d m_1 f_{\rm IMF}(m_1) }\ .
\end{equation}
Here, $f_q(q)$ is the mass ratio distribution. We find $\eta=0.21$ and $\eta=0.25$, for an uniform and log-uniform mass ratio distributions, respectively. A factor of uncertainty is the possible KL dynamics during the evolution of the stellar triples before they form a SBH-MS system in the inner binary, which we have not modeled here. Some fraction of the parameter space can be removed by the earlier evolution of the system \citep{shapp2013}. To estimate this uncertainty, we consider very conservatively that any stellar triple whose initial KL timescale is less than the lifetime of the primary star \citep[$\sim 7$ Myr;][]{iben91,hurley00,maeder09} in the inner binary merges in the main sequence phase, and, as a consequence, will not form a triple system with an inner SBH-MS binary \citep{rodant2018}. We find that the fraction of these triples is $\zeta\sim 0.8$ on average, except for Model A2 where we find $\zeta\sim 0.7$.

In our calculations, we adopt $f_{\rm 3}=0.25$. The fraction of stable systems after SNe depends mainly on the value of $\sigma$ for the natal velocity kick distribution. We find $f_{\rm stable}\approx 9.2\times 10^{-2}$, $8.1\times 10^{-3}$, $1.6\times 10^{-3}$ for $\sigma=0\kms$, $13\kms$, $34\kms$, respectively, when $\amax=2000$ AU, and $f_{\rm stable}\approx 1.6\times 10^{-3}$, $1.1\times 10^{-3}$, $1.0\times 10^{-3}$ for $\amax=2000$ AU, $5000$ AU, $7000$ AU, respectively, when $\sigma=260\kms$. The typical fraction of systems that produce a TDE is $f_{\rm TDE}=0.13$ (see Tab.~\ref{tab:models}). Here, we include also systems that result in crossing of the Roche limit. This can  produce phases of accretion as X-ray binaries and may drive the two stars into merging \citep[e.g.][]{naozfab2014}. We find that typically these systems constitute $\sim 30\%$--$50\%$ of the merging systems. Using the minimum and maximum values of $f_{\rm stable}$ in Tab.~\ref{tab:models}, our final estimated rate is in the range,
\begin{equation}
\mathcal{R}_\mathrm{TDE}=0.11-16 \ \mathrm{Gpc}^{-3}\ \mathrm{yr}^{-1}\ .
\end{equation}
For a log-uniform distribution of mass ratios, we estimate a rate $\sim 2$ times larger. Considering the signal up to $z=0.1$, the TDE rate becomes,
\begin{equation}\label{eq:Gamma1}
\Gamma_\mathrm{TDE}(z\le 0.1)=3.4\times 10^{-2}-4.7 \ \mathrm{yr}^{-1}\ .
\end{equation}
We have also computed the values of $f_{\rm stable}$ in the case of non-momentum-conserving kicks, i.e. all the SBHs (and NSs) receive a kick independent of their mass \citep[see e.g.][]{pernaccb2018}. We have found that $f_{\rm stable}\sim 9.2\times 10^{-2}$, $2.2\times 10^{-4}$, $2.6\times 10^{-5}$, $1.7\times 10^{-6}$ for $\sigma=0\kms$, $\sigma=50\kms$, $100\kms$, $260\kms$, respectively. Thus, assuming an average $f_{\rm TDE}=0.15$, the predicted local TDE rate is for non-momentum-conserving kicks,
\begin{equation}\label{eq:Gamma2}
\Gamma_\mathrm{TDE,3}^{\rm NC}(z\le 0.1)=2.1\times 10^{-4}-4.7 \ \mathrm{yr}^{-1}\ .
\end{equation}

Finally, we can estimate the TDE rate in triples for a Milky Way-like galaxy. Assuming momentum-conserving natal kicks and a star formation rate of $\sim 1\msun\ \mathrm{yr}^{-1}$ \citep{licq2015},
\begin{equation}\label{eq:Gammamw}
\Gamma_\mathrm{TDE}^\mathrm{MW}=1.3\times 10^{-9}-1.9\times 10^{-7} \ \mathrm{yr}^{-1}\ .
\end{equation}
For comparison, the rates for SMBHs and IMBHs are estimated to be $10^{-5}-10^{-4}$ yr$^{-1}$ \citep{vanvelzen2018} and $10^{-5}-10^{-3}$ yr$^{-1}$ \citep{fraglgk2018}, respectively. 

We note that we are not taking into consideration fallback, whose effect would likely increase the TDE rates for large $\sigbh$'s.

\section{Electromagnetic signatures of stellar Tidal Disruption Events}
\label{sect:em}

How can these events be recognized among the variety of transient sources on the sky? TDEs onto massive BHs (MBHs) ubiquitous in galactic centers have long been studied in the literature, partly motivated by actual detections of TDEs by MBHs. However, the basic physics of the phenomenon is expected to be similar across different scales, with the mass of the disrupting BH mostly dictating the characteristic accretion rate (and hence luminosity) and the duration of the event.

Before discussing the above, we need to assess whether the presence of the outer companion is expected to be influential on the TDE event. In fact, it has been discussed in the context of MBHs that binarity can impact the rate of mass fallback and the formation of an accretion flow following the disruption event \citep{Liu2014,Coughlin2017}. In particular, fallback dynamics will be drastically perturbed if the apocenter distance $a_{\rm apo}$ of fallback orbits is such that $a_{\rm apo}>\aout$.  This translates into $\aout <  3\times 10^{-6}(\mms/\msun)^{1/3} (T/1\,{\rm month})^{2/3}$~pc, where $T$ is the time after disruption. Comparing this distance with the distribution of separations of the outer companion of the (former) triple studied here (cf. Fig.~\ref{fig:ainaout}), we see that $a_{\rm out}\gg a_{\rm apo}$, and hence the presence of the outer companion is not expected to affect the fallback dynamics.

Once the star (of mass $\mms$ and radius $R_*$) has been disrupted, elements of the tidal debris move on nearly geodesic orbits around the BH. In order to circularize, i.e. to form an accretion disk, the bound matter must lose a significant amount of energy. Circularization is believed to be possible thanks to general relativistic effects, as apsidal precession forces highly eccentric debris streams to self-intersect (i.e. \citealt{stone2019}). However, whether circularization can be completed before the end of the actual event still remains an issue of debate \citep{Piran2015}; albeit, in the case of SBHs, it is aided by the fact that the bound debris are not highly eccentric \citep{Kremer2019}. A large fraction of the debris is expected to be flung out and become unbound as a result of heating associated with inter-stream shocks \citep{Ayal2000}. For the material that remains bound, the fallback rate after an initial rapid increase follows roughly the decay \citep{phinn89}
\begin{equation}
\dot{m}_{\rm MS}\sim \frac{\mms}{ t_0}\left(\frac{t}{t_0}\right)^{-5/3}\,,
\label{mdot}
\end{equation}
where $t_0$ represents the time after which the first bound material
returns to pericenter ($R_p$), with $R_p\sim R_T$, (see e.g. \citealt{Guillochon2013,Stone2013,Stone2019a}), i.e.
\begin{eqnarray}
t_0 &=& \frac{\pi R^3_T}{\sqrt{2G\mbh R_*^3}} \approx\nonumber\\
&\approx & 9\times 10^3{\rm s} \left(\frac{R_T}{\rsun}\right)^3\left(\frac{\rsun}{R_*}\right)^{3/2}\left(\frac{10\msun}{\mbh}\right)^{1/2}\nonumber \\
&=&9\times 10^3{\rm s} \left(\frac{R_*}{\rsun}\right)^{3/2}\left(\frac{\mms}{\msun}\right)^{-1} \left(\frac{\mbh}{10\msun}\right)^{1/2}\,.
\label{eq:t0}
\end{eqnarray}
The late time fallback lightcurve is often observed to be shallower than $t^{-1}$ \citep{Auchettl2017} which may be due to general relativistic corrections to viscous diffusion and finite stress at the last stable orbit \citep{balbus2018}. Note however that, in the cases for which only partial disruption of the star is achieved, then the fallback rate is found to be steeper than $t^{-5/3}$ (e.g. \citealt{Guillochon2013}). Recently, \citet{coug2019} showed that the fallback rate from a partial TDE follows a universal scaling of $t^{-9/4}$.

In our calculations, we are assuming that a star is disrupted whenever its closest passage to the SBH is smaller than $R_T$ (Eq.~\ref{eqn:rtid}). We note that $R_T$ is only the leading order term in the ratio $R_*/r_{\rm CM,*}$, where $r_{\rm CM,*}$ is the position of the center of mass of the star. Higher-order terms in the gravitational potential (as the binding energy of the disrupted star itself) could remain important if comparable to the spread induced by the SBH tidal potential, and this will further modify the energy distribution of the disrupted debris. As a consequence, a larger fraction of the debris produced by the TDE may be bound to the SBH, resulting in a larger peak accretion rate and corresponding luminosity.

Once a disk is formed, the timescale for the tidally disrupted debris to accrete is set by the viscous timescale. For a geometrically thick disk (i.e with a scale height $H/R\sim 1$), as expected
at high accretion rates, this is given by
\begin{equation}
t_{\rm acc}\sim \frac{1}{\alpha \Omega_{\rm K}(R_{\rm in})}
=4\times 10^4\,{\rm s} \left(\frac{\alpha}{0.1}\right)^{-1}
\left(\frac{R_{in}}{2\rsun}\right)^{3/2}
\left(\frac{10\msun}{\mbh}\right)^{1/2}\ ,
\label{eq:tvisc}
\end{equation}
where $\Omega_{\rm K} (R_{\rm in})$ is the Keplerian angular speed at the radius of closest approach $R_{\rm in}=R_p$, and $\alpha$ the viscosity constant \citep{Shakura1973}. The maximum accretion rate, $\dot{M}_{\rm acc}\sim M_{\rm fb}/t_{\rm acc}$, with $M_{\rm fb}$ being the fraction of $\mms$ which remains bound and falls back, is found to be in the range of $\sim$ few $\times 10^{-6}$ - few  $\times 10^{-5}$ $\msun$~s$^{-1}$ \citep{Perets2016,Kremer2019}, and it declines at later times as a power-law, as discussed above. These accretion rates are somewhat lower, but still relatively comparable, to the fallback rates found in numerical simulations of the collapse of blue supergiant stars \citep{Perna2018b}, which have been invoked as possible progenitors of the small but interesting sub-class of ultra-long GRBs \citep{Gendre2013}. If a jet is driven as observed in TDEs from MBHs, then the phenomenology may indeed appear similar to that of the ultra-long GRBs, as pointed out by \citet{Perets2016}, i.e. a bright and energetic flare in $\gamma$-rays and X-rays, with durations on the order of $10^{3}-10^{4}$~s. These events would be distinguishable from the ultra-long GRBs, in that they would not have a supernova associated with them. Furthermore, the TDEs formed in stellar mass triples are expected to have different spectra, since they are not limited to any particular stellar type such as Wolf-Rayet or blue supergiants which produce long and ultra long GRBs, respectively. 

The above scenario assumes that a jet is successfully launched. However, this may not necessarily be the case: the physics of hyper-Eddington accretion is still not fully understood (see e.g. \citealt{Abramowicz2013}), and neither are the conditions required for the production of a relativistic, 'GRB-like' jet. Several studies (i.e. \citealt{Narayan1994,Blandford1999b}) have shown that strong winds can blow away a significant fraction of the disk mass. The radiation-hydrodynamic evolution of such a wind was recently calculated by \citet{Kremer2019}; they found that it can give rise to optical transients with luminosities $\sim 10^{41}-10^{44}$~erg~s$^{-1}$, lasting from a day to about a month. An event with a typical luminosity of $10^{42}$~erg~s$^{-1}$ would be observable with the Zwicky Transient Facility (ZTF) up to a distance of $\sim 150$~Mpc.
The higher sensitivity of the upcoming LSST survey  ($r < 24.5$ for 15~s of integration time during the routine sky scans) would allow detection up to about 0.5~Gpc for a luminosity of $\sim 10^{41}$~erg~s$^{-1}$.

In the most pessimistic scenario, in which a jet is not	launched and
the unbound material is not strongly emitting, the dominant source of  luminosity
would be provided by the accretion disk (such as in
advection-dominated accretion disks with super-Eddington accretion and low radiative efficiency,
e.g. \citealt{Popham1999,Narayan2001,Janiuk2004}) and
it would be  Eddington-limited, hence not expected to exceed $\sim 10^{40}$~ergs. These events, mostly brighter in X-rays, would constitute the high-end of the high-mass X-ray binary luminosity function, but would be transient.

Additional emission could arise even \textit{prior} to the TDE event as
the gas liberated by the SN could accrete onto the SBH, initiating X-ray emission. These systems would look like high-mass X-ray binaries with a steady increase in X-rays, which have a sudden jump in brightness in X-rays, and/or at other wavelengths as a consequence of the TDE, if the latter takes place not too long after the SN explosion.

\section{Discussion and conclusions}
\label{sect:conc}

In this paper we have investigated the possibility that triple stars are sources of TDEs, some of which can take place off-center from the nucleus of the host galaxy. We started from a triple system made up of three MS stars and modeled the SN event that leads to the production of an inner binary comprised of a SBH. We evolved these triples with a high precision $N$-body code and studied their TDEs as a result of KL oscillations. We adopted different distributions of natal kicks imparted during the SN event, different maximum initial separations for the triples, and different distributions of eccentricities. Most of the systems produce a TDE when the relative inclination of the inner and outer orbits is $\sim 90^\circ$ after the SN event takes place. We showed that the main parameter that governs the properties of the SBH-MS binaries that produce a TDE in triples is the mean natal velocity kick. Smaller values lead to larger inner and outer semi-major axes of the systems that undergo a TDE, smaller SBH masses, and larger merging timescales.

While the fraction of systems that remain in a stable triple $f_{\rm stable}$ depends critically on the prescriptions adopted for natal kicks, we found that the fraction of these systems that produce a TDE, $f_{\rm TDE}$, is not significantly affected by the initial conditions considered in this work and is typically $\sim 15\%$ (see Tab.~\ref{tab:models}). Therefore, the future observed TDE rates of stars onto SBHs could be used to constrain the underlying kick distribution. We have estimated the rate of TDEs in triples in the range $3.4\times 10^{-2}-4.7 \ \mathrm{yr}^{-1}$ for $z\leq 0.1$, assuming momentum-conserving natal kicks, consistent with the recent rate inferred for star clusters by \citet{Kremer2019}. For a Milky Way-like galaxy, we found a rate in the range $1.3\times 10^{-9}-1.9\times 10^{-7} \ \mathrm{yr}^{-1}$. As a comparative reference, the rates for SMBHs and IMBHs are estimated to be $10^{-5}-10^{-4}$ yr$^{-1}$ \citep{vanvelzen2018} and $10^{-5}-10^{-3}$ yr$^{-1}$ \citep{fraglgk2018}, respectively. If the MS stars evolves to form a WD, SBH-WD binaries can merge under the influence of the third companion via KL cyles and are expected to generate other kinds of transients \citep{fetal19}. 

We used simple prescriptions to determine the final NS and SBH masses, without entering into  the details of the single and binary stellar evolution, and hence the dependence of the compact remnants on metallicity, stellar winds, rotation, Roche lobe overflows and possible common envelope phases. These effects could reduce the parameter space \citep{shapp2013} studied here; however, they still lack a full,  self-consistent description in triple systems, due to the possible interplay with the KL cycles \citep{rosa2019,hamd2019}. This could have an effect on the final rates estimated in this work. We leave it to a future study to quantify the importance of these effects in the context of triple stellar evolution.

A triple system comprised of an inner SBH-MS binary orbiting an SMBH could similarly produce TDEs onto SBHs. The triple dynamics would be similar, but would take place on a shorter timescale owing to the large SMBH mass. The typical rate of TDEs coming from this channel would then depend on the supply rate of binaries that produce an SBH-MS binary. This could be due to a slow segregation of binaries from the outermost regions of the SMBH sphere of influence or local star formation \citep{fragrish2018}. Thus, if there is a nuclear sturbust that would populate the innermost regions of the galaxy with these kind of binaries, a burst of TDEs onto SBHs is expected coming from the centre of the galaxy. Therefore, these TDEs would come from the galactic nucleus as the usual TDEs onto SMBH.

Detectability of these events will depend on the dominant emission mechanism. If a jet is successfully launched, then the observable phenomenology is going to resemble that of an ultralong GRB. So far, only a handful of these has been observed. Hence the upper bound of our predicted rates is already starting to be probed by the rates of these events, at least for jets with a wide angular size.  
Additional (or alternative emission of no jet is successfully  launched) could come from a disk-wind \citep{Kremer2019}, and would be peaking in the optical; for typical luminosities it would be detectable by LSST with up to a few events per year.
In the most pessimistic scenario in which neither jets nor winds are launched, these systems would populate the tail of the high-mass X-ray binaries, but would be transient; some X-ray emission could be seen even preceding the TDE.

\section*{Acknowledgements}

We thank the anonymous referee for constructive comments. GF thanks Seppo Mikkola and Fabio Antonini for helpful discussions on the use of the code \textsc{archain}. This work received funding from the European Research Council (ERC) under the European Union’s Horizon 2020 Programme for Research and Innovation ERC-2014-STG under grant agreement No. 638435 (GalNUC) and from the Hungarian National Research, Development, and Innovation Office under grant NKFIH KH-125675 (to BK). GF is in part supported by the Foreign Postdoctoral Fellowship Program of the Israel Academy of Sciences and Humanities. This research was supported in part by the National Science Foundation under Grant No. NSF PHY-1748958. RP acknowledges support by NSF award AST-1616157. The Center for Computational Astrophysics at the Flatiron Institute is supported by the Simons Foundation. Simulations were run on the \textit{Astric} cluster at the Hebrew University of Jerusalem.

\bibliographystyle{mn2e}
\bibliography{refs}

\end{document}